\documentclass{article}
\usepackage{bm, amsmath, amsfonts, amssymb}
\usepackage{color}
\usepackage{graphicx}
\usepackage{comment}

\begin{document}

% Page header
%\markboth{Okuma \bullet \ Sato}{Non-Hermitian topological phenomena}

% Title
\title{\textbf{Non-Hermitian topological phenomena: A review}}

%Authors, affiliations address.
\author{Nobuyuki Okuma$^1$ and Masatoshi Sato$^1$
%\affil{$^1$Center for Gravitational Physics and Quantum Information, Yukawa Institute for Theoretical Physics, Kyoto University, Kyoto 606-8502, Japan; email: okuma@hosi.phys.s.u-tokyo.ac.jp, msato@yukawa.kyoto-u.ac.jp}
}

%Abstract
\date{}

%Keywords, etc.
%\begin{keywords}
%Non-Hermitian physics, topological phase, bulk-boundary correspondence, spectral theory
%\end{keywords}
\maketitle

%Table of Contents
%\tableofcontents

%%%%%%%%%%
% main text
\noindent{\footnotesize$^1$Center for Gravitational Physics and Quantum Information,\\
Yukawa Institute for Theoretical Physics, Kyoto University, Kyoto 606-8502, Japan;\\
email: okuma@hosi.phys.s.u-tokyo.ac.jp, msato@yukawa.kyoto-u.ac.jp}
\begin{abstract}
The past decades have witnessed an explosion of interest in topological materials, and a lot of mathematical concepts have been introduced in condensed matter physics. Among them, the bulk-boundary correspondence is the central topic in topological physics, which has inspired researchers to focus on boundary physics. Recently, the concepts of topological phases have been extended to non-Hermitian Hamiltonians, whose eigenvalues can be complex. Besides the topology, non-Hermiticity can also cause a boundary phenomenon called the non-Hermitian skin effect, which is an extreme sensitivity of the spectrum to the boundary condition. In this article, we review developments in non-Hermitian topological physics by focusing mainly on the boundary problem. As well as the competition between non-Hermitian and topological boundary phenomena, we discuss the topological nature inherent in non-Hermiticity itself. 
\end{abstract}

\section{INTRODUCTION}
Recently, the topological properties of lattice systems have attracted broad interest in condensed matter physics.
One of the major concepts in topological physics is the bulk-boundary correspondence in topological phases \cite{Kane-review,Zhang-review}, which states that the topological invariant constructed from the bulk states counts the number of robust gapless boundary states \cite{Hatsugai-93}.
In this context, topological phenomena also have aspects as boundary phenomena.

Dynamical properties of open systems are other major perspectives in recent condensed matter physics.
While the Hamiltonian represented by a Hermitian matrix is the central object of interest in isolated equilibrium systems, non-Hermitian matrices, whose spectrum can be complex, play important roles in various open systems.
Although the absence of Hermiticity brings about troublesome properties that defy the common sense of condensed matter physicists, it leads to new and deeper physics as well. In recent years, a lot of concepts in spectral theory \cite{Trefethen} and non-Hermitian physics \cite{Bender-98, Bender-02, Bender-review, Konotop-review, Christodoulides-review} have been introduced in condensed matter physics \cite{ashida-gong-20}.

In this paper, we review the topological physics in non-Hermitian lattice systems by focusing mainly on the boundary phenomena.
The concept of topological phases was generalized to non-Hermitian systems in Refs. \cite{Rudner-09, Hu-11, Esaki-11,Schomerus-13, Shen-18}. 
In particular, Ref. \cite{Esaki-11} provided bulk topological numbers in various non-Hermitian systems and showed the validity of the bulk-boundary correspondence in the presence of a class of symmetries.  
However, it was pointed out later that a general non-Hermitian system without such symmetries may exhibit other boundary phenomena specific to non-Hermitian systems \cite{YW-18-SSH,Kunst-18}, called non-Hermitian skin effects \cite{YW-18-SSH}.
The new effects localize bulk modes on boundaries and obscure the bulk-boundary correspondence.
We first review the competition between the bulk-boundary correspondence and the non-Hermitian skin effects, 
but our focus is not merely on the competition between the two different phenomena.
Remarkably, the latter non-Hermitian boundary physics is also closely related to topological physics.
In the main part, we introduce the mathematics of non-Hermitian matrices and discuss recent developments in the field of non-Hermitian phenomena whose origin is {\it non-Hermitian topology}.

\section{REVIEW OF NON-HERMITIAN SKIN EFFECT}
\subsection{Example and definition of non-Hermitian skin effect}
In conventional solid-state physics, boundary insensitivity of bulk quantities plays important roles because one can freely choose a convenient boundary condition.
In particular, the periodic boundary condition (PBC), together with translation invariance in solids, enables us to introduce band theory based on the momentum-space picture. 

\begin{figure}
\begin{center}
 \includegraphics[width=12cm,angle=0,clip]{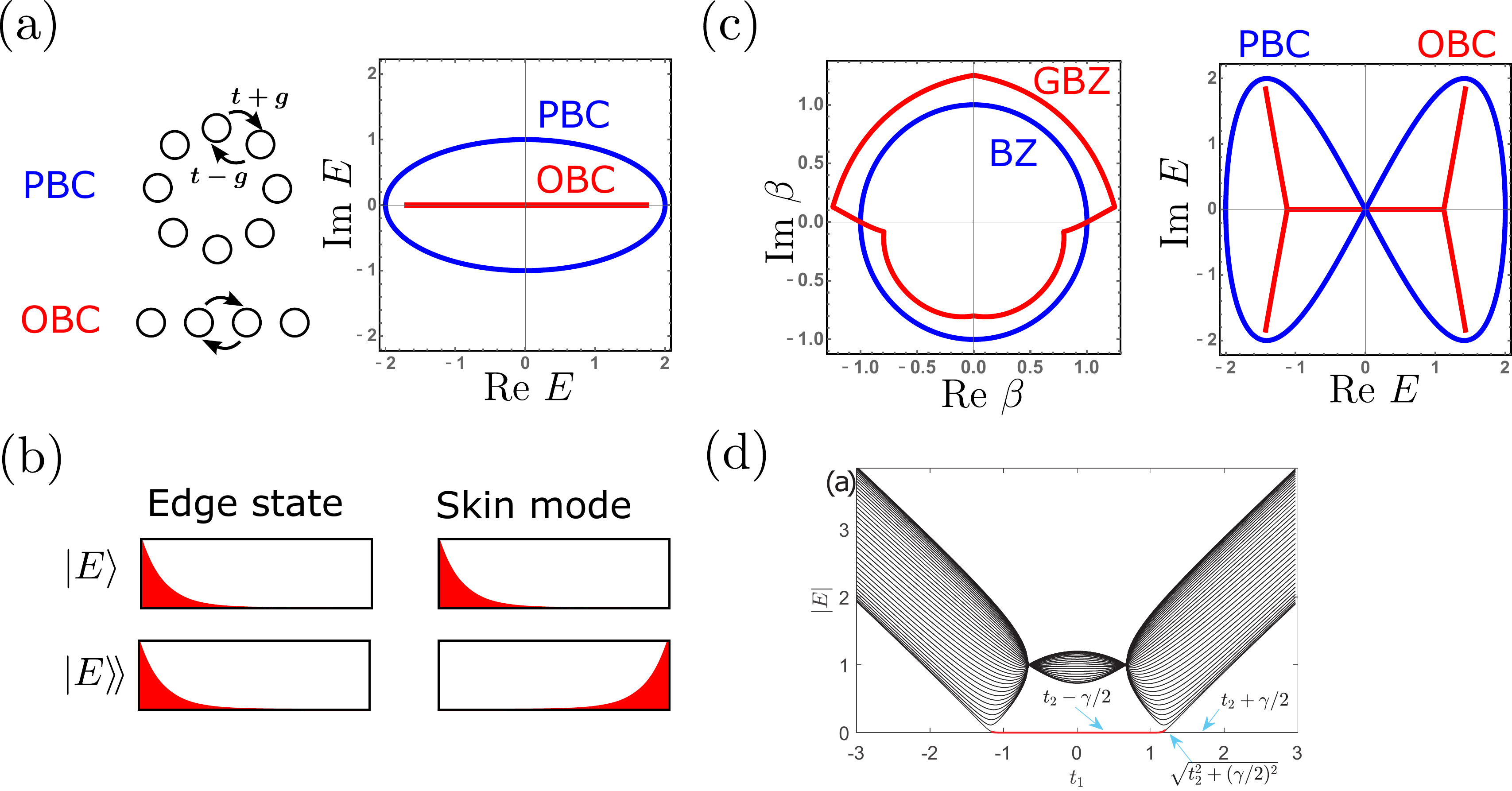}
 \caption{(a) The PBC and OBC eigenspectrum of Hatano-Nelson model ($t=1$,$g=0.5$). (b) Right and left eigenstates of a Hermitian edge state and a non-Hermitian skin mode. (c) Brilloin zone (BZ) and generalized Brillouin zone (GBZ) for PBC and OBC spectra, respectively. We use $E(\beta)=\beta^2-i\beta+i\beta^{-1}-\beta^{-2}$ \cite{Trefethen,Okuma-correlated}.  (d) The OBC spectrum (absolute value) of the non-Hermitian Su-Schrieffer-Heeger model as a function of model parameter, adopted from Ref. \cite{YW-18-SSH}. The OBC and PBC gap-closing points are indicated by arrows. The emergence of topological zero modes (red) under the OBC cannot be predicted by the PBC information. }
 \label{fig1}
\end{center}
\end{figure}

In non-Hermitian physics, on the other hand, the boundary condition can drastically affect bulk properties.
Such a boundary sensitivity is best represented by the spectral properties of the Hatano-Nelson model (without disorder), which is a non-Hermitian tight-binding Hamiltonian defined on a one-dimensional lattice
\cite{Hatano-Nelson-96,Hatano-Nelson-97, Hatano98}:
\begin{align}
    \hat{H}_{\rm HN}=\sum_{i}\left[(t+g)c^{\dagger}_{i+1}c_i+(t-g)c^{\dagger}_i c_{i+1}\right],\label{HNmodel}
\end{align}
where $i$ is the site index, $(c^{\dagger},c)$ are the bosonic or fermionic creation and annihilation operators, and $t\in\mathbb{R}$ and $g\in\mathbb{R}$ represent the Hermitian symmetric and non-Hermitian asymmetric hopping terms, respectively. 
We express the Hamiltonian in the equivalent matrix representation henceforth:
\begin{align}
H_{\mathrm{HN}}:=
\begin{pmatrix}
0&t-g&0&\cdots\\
t+g&0&t-g&\cdots\\
0&t+g&0&\cdots\\
\vdots&\vdots&\vdots&\ddots
\end{pmatrix}.
\end{align}
Under the PBC $([H_{\mathrm{HN}}]_{1,N/N,1}=t\pm g)$, we obtain the complex energy spectrum of the Hamiltonian by using the Fourier transform: $E_{k}=(t+g)e^{ik}+(t-g)e^{-ik}$, where $k=2\pi j/L$ ( $j=0,1,\cdots,L-1$) with $L$ being the number of sites.
For $tg\neq0$, the PBC spectrum is on an ellipse in the complex plane.
Under the open boundary condition (OBC), on the other hand, the eigenspectrum is completely different:
Using the imaginary gauge transformation\footnote{This term was named after the fact that this transformation can be interpreted as a gauge transformation with an imaginary phase $\theta$:  $c_j\rightarrow e^{i\theta j}c_j,~c^{\dagger}_j\rightarrow e^{-i\theta j}c^{\dagger}_j$. }, one can map the OBC Hamiltonian to a Hermitian Hamiltonian without asymmetry \cite{Hatano-Nelson-96,Hatano-Nelson-97}:
\begin{align}
H^{(\rm sim)}_{\rm HN}&:= V_{r}^{-1} H_{\rm HN} V_{r}\notag\\
&=
\begin{pmatrix}
0&\sqrt{t^2-g^2}&\cdots\\
\sqrt{t^2-g^2}&0&\cdots\\
\vdots&\vdots&\ddots
\end{pmatrix},\label{imggauge}
\end{align}
where $[V_r]_{i,j}=r^i\delta_{ij}$ with $r=\sqrt{t+g/t-g}$, and 
we have assumed $t>g>0$ for simplicity.
Since this is a similarity transformation, preserving the eigenspectrum of a finite-dimensional matrix, the OBC spectrum of the Hatano-Nelson model is given by that of the Hermitian matrix $H^{(\rm sim)} _{\rm HN}$. 
Thus, the OBC spectrum is on a line on the real axis of the complex plane, 
which is very different from the PBC one on an ellipse [Figure \ref{fig1}(a)].
The corresponding OBC eigenstates, which are obtained from those of $H^{(\rm sim)} _{\rm HN}$ by the similarity transformation $V_r$, are exponentially localized at a boundary, in contrast to the extended PBC eigenstates.

Recently, such extreme sensitivities of eigenspectra and eigenstates against the boundary condition have been extensively studied as the {\it non-Hermitian skin effect}, named by Yao and Wang \cite{YW-18-SSH}. In their work, this term was used in the following situation \cite{YW-18-SSH}:
\\
\\
{\it all the eigenstates of an open chain are found to be localized near the boundary} 
\\
\\
In this paper, we adopt a broader definition of the non-Hermitian skin effect.
We focus on the boundary-localized modes whose origin is non-Hermiticity.
Unlike eigenstates of Hermitian Hamiltonians, the Hermitian conjugate of a right eigenstate (ket eigenvector) is not always a left eigenstate (bra eigenvector) in the presence of non-Hermiticity\footnote{More precisely, this is property of non-normal matrices discussed later.}. To emphasize the difference, we adopt the following notation:
\begin{align}
H|E\rangle=E|E\rangle,~
    \langle\!\langle E|H=E\langle\!\langle E|.\label{rightleft}
\end{align}
We call a mode $|E\rangle$ as a {\it non-Hermitian skin mode} if its spatial distribution is different from that of the 
left counterpart $|E\rangle\!\rangle$.
Typically, we are interested in skin modes localized at one boundary whose left counterparts are localized at the other boundary [Figure \ref{fig1}(b)].
Such a situation is realized in the case of the Hatano-Nelson model under the OBC.
In the following, we define the non-Hermitian skin effect by the presence of non-Hermitian skin modes. As we will see throughout the paper, the presence of skin modes is equivalent to the extreme sensitivity of the non-Hermitian energy spectrum to the boundary condition.

In the first half of this paper, we will focus on {\it conventional} skin effects with $\mathcal{O}(L)$ skin modes\footnote{We count the number of modes by distinguishing algebraic degeneracy (multiplicity) for an eigenvalue $E$ defined as the number of times $\lambda=E$ appears as a root of $\det (H-\lambda)$. For example, there is only one eigenvector for Eq.(\ref{HNmodel}) with $t=g$ because of the nondiagonalizability.
In this case, the number of skin modes is counted as $L$, while the geometric degeneracy (multiplicity) defined as the dimension of the eigenspace is one.   }, which are found in one-dimensional systems with discrete translation invariance (except for the ends).
In the presence of the conventional skin effect, the OBC bulk spectrum is determined by the non-Bloch band theory \cite{Schmidt-Spitzer-60,YW-18-SSH,YSW-18-Chern,Kunst-18,YM-19,yokomizo2020non,Yang-19,KOS-20}.
In conventional band theory, the bulk spectrum, or the energy dispersion, is given as a function of crystal momentum $k$ (or the exponential factor of a Bloch wavefunction $e^{ik}$), $E(e^{ik})$. 
On the other hand, the non-Bloch band theory provides the OBC bulk spectrum as an analytic continuation $E(e^{ik})\rightarrow E(\beta)$, where $\beta$ is a complex parameter on a closed curve in the complex plane, called the generalized Brillouin zone (GBZ).
In the absence of the non-Hermitian skin effect, the GBZ reproduces the unit circle of the BZ, while in the case of the Hatano-Nelson model, it is given by a circle with radius $r^{-1}$.
In general, the GBZ can be a complicated curve [Figure \ref{fig1}(c)] and is determined with the help of numerical calculations, as discussed in Refs. \cite{YM-19, Yang-19} \footnote{The factors $e^{ik}(\beta)$ in our notation correspond to the factors $e^{-ik}(\beta^{-1})$ in Ref. \cite{YM-19}.}.

\subsection{Bulk-edge correspondence under non-Hermitian skin effect\label{breakdown-of-bbc}}
The non-Hermitian skin effect obscures the bulk-boundary correspondence \cite{Lee-16,MartinezAlvarez-18,Xiong-2018,Kunst-18,YW-18-SSH,YSW-18-Chern}. More precisely, even when the PBC spectrum is gapless and thus one cannot define the bulk topological invariant in the PBC, the bulk OBC spectrum can be gapped due to the skin effect.  
As a result, the topological phase diagram under the OBC is generally different from that under the PBC.
In Figure \ref{fig1}(d), the OBC spectrum of a non-Hermitian extension of the Su-Schrieffer-Heeger model \cite{SSH-79} with asymmetric hopping \cite{Lee-16} is plotted as a function of a model parameter \cite{YW-18-SSH}. This figure shows that the topological phase transition with the appearance of topological zero modes occurs at a point that is different from the gap-closing points under the PBC.

At first sight, this phenomenon suggests a breakdown of the bulk-boundary correspondence.
However, even in such a case, one can recover the bulk-boundary correspondence once one introduces a proper topological invariant via the OBC bulk spectral information.
For example, the topological invariant for the non-Hermitian Su-Schrieffer-Heeger model is given by a winding number defined in terms of the GBZ \cite{YW-18-SSH} in the OBC, instead of the conventional BZ in the PBC.
To treat more generic cases including higher-dimensional systems and disordered ones,
Ref. \cite{song-real-space-19} proposed to use real-space topological invariants in the OBC, which enables us to characterize the bulk-boundary correspondence in terms of the explicit right and left eigenstates of the OBC Hamiltonian.
These studies indicate that the bulk-boundary correspondence should be defined for the bulk and boundary modes under the common boundary condition. In the absence of the non-Hermitian skin effect, on the other hand, the topological invariant under the OBC coincides with that under the PBC.
In this sense, the success of predictions by the PBC information in Hermitian topological physics is due to not only the bulk-boundary correspondence but also the boundary insensitivity, which is a special property of the systems without skin modes.
Such a conventional bulk-boundary correspondence also holds in non-Hermitian systems without skin effects.
Actually, when a class of symmetries may prohibit the skin effects \cite{KSUS-19}, the topological invariant under the PBC succeeds in the prediction of the boundary modes in the OBC \cite{Esaki-11}.
One can also obtain the conventional bulk-boundary correspondence for the bosonic Bogoliubov-de Gennes Hamiltonian \cite{bogoliubov1947theory,Altland-Simons,Colpa-78,kawaguchi2012spinor, Shindou-13}, whose energy spectrum is given by the eigenspectrum of a non-Hermitian matrix with pseudo-Hermiticity and particle-hole symmetry \cite{BdGsym, KSUS-19} \footnote{For more details, see a review paper of topological magnons \cite{topo-magnon}.}.
It should be noted here that one needs to generalize the topological invariants properly so as to be consistent with non-Hermiticity \cite{Esaki-11, Shen-18, Shindou-13,Ghatak-2019} even without the skin effects.

In this subsection, we have seen the competition between two different boundary phenomena: the bulk-boundary correspondence and the non-Hermitian skin effect.
Actually, the non-Hermitian skin effect itself is also a kind of topological phenomenon, as we shall see in the following sections.

\section{Conventional skin effect as non-Hermitian topological phenomenon}
In this section, we review the topological nature of the conventional non-Hermitian skin effect \cite{OKSS-20,Zhang-19}.
In particular, we focus on the fact that the conventional non-Hermitian skin effect and the bulk-boundary correspondence in a class-AIII one-dimensional topological insulator share the same mathematical origin.

\subsection{Winding number as an indicator for conventional skin effect}
As we mentioned, the OBC spectrum is given by a function of $\beta$ on the GBZ determined by the non-Bloch band theory.
Unfortunately, the shape of the GBZ is not a simple function of the model parameter in general, and the spectral behavior is unclear without the help of numerical calculations. 
Nevertheless, the presence or absence of the non-Hermitian skin effect is easily determined by the $non$-$Hermitian$ $topology$, i.e., the winding number $W\in\mathbb{Z}$ of the PBC spectral curve in the complex plane.
Several observations about this correspondence were indicated in Refs. \cite{Gong-18, Lee-Thomale-19, Borgnia-19}, and related theorems were proven in Refs. \cite{OKSS-20,Zhang-19}.
In the following, we focus on the theorem in Ref. \cite{OKSS-20}.
Suppose that the system is described by a one-dimensional translation-invariant non-Hermitian tight-binding model $H$ with finite-range hopping and without symmetry.
Then the following theorem holds in the infinite-volume limit \cite{OKSS-20}\footnote{For a rigorous definition of the infinite-volume limit, see the math textbook \cite{Bottcher}.  }. 
\\

{\bf Theorem}~~The OBC bulk spectrum cannot have a non-trivial winding number.
Consequently, the PBC spectral curve with a non-trivial winding number implies that the OBC bulk spectral curve is different from the PBC one, or equivalently, the conventional skin effect inevitably occurs.\\

\begin{figure}
\begin{center}
 \includegraphics[width=12cm,angle=0,clip]{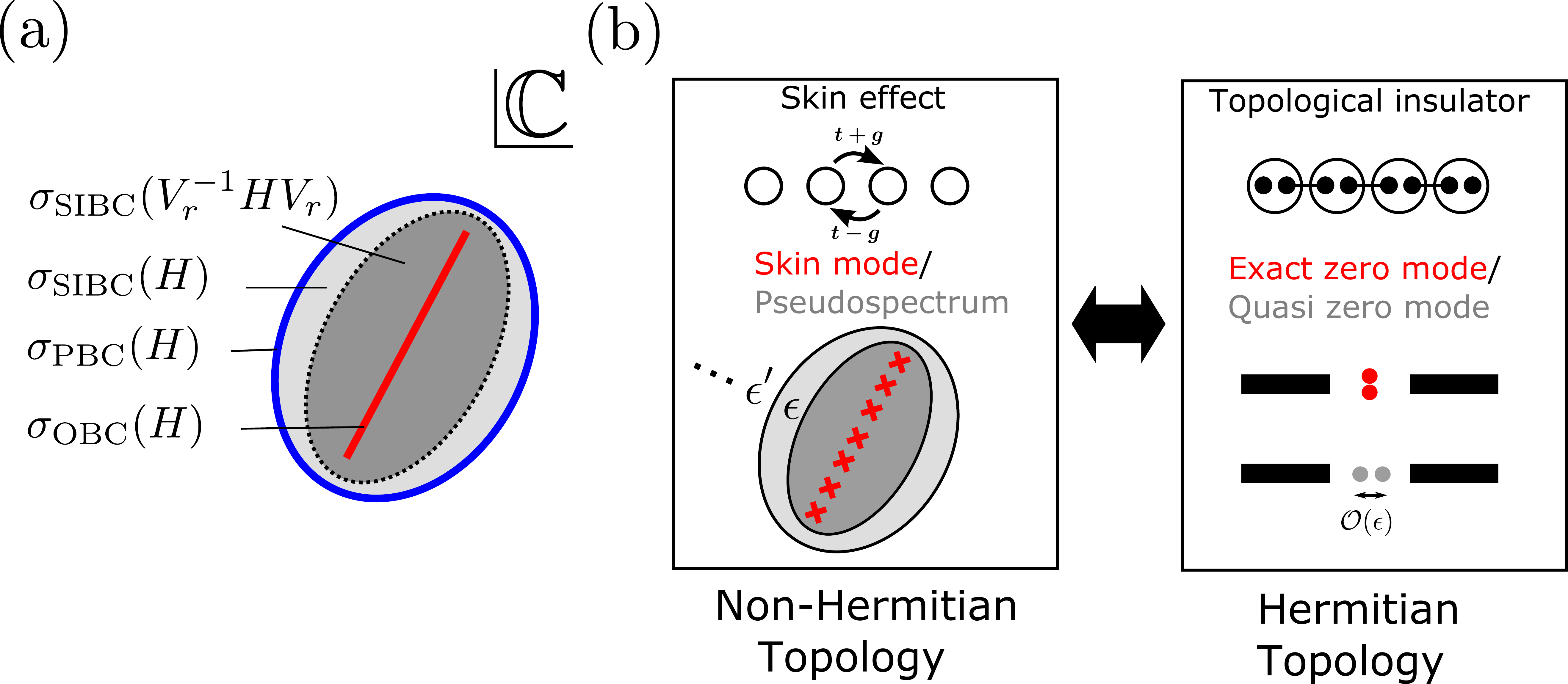}
 \caption{(a) Schematic picture of the PBC, SIBC, and OBC spectra in the presence of non-Hermitian skin effect. The disk with the dotted boundary is the SIBC spectrum under an imaginary gauge transformation $V_r$. (b) Relationship between non-Hermitian topology and Hermitian topology. }
 \label{fig2}
\end{center}
\end{figure}

We here describe a strategy for the proof.
Since there is no simple formula for calculating the OBC spectrum,
it is difficult to compare the OBC spectrum to the PBC one directly. Thus, the proof of this theorem is divided into two parts: (i) the comparison between the PBC and the {\it semi-infinite} boundary condition (SIBC), and (ii) that between the SIBC and the OBC.
Here the ``semi-infinite" means that there is only one boundary in the one-dimensional lattice.
In step (i), we use the index theorem of spectral theory \cite{Trefethen,Bottcher} to relate the SIBC spectrum with the PBC one.
According to the index theorem, the SIBC spectrum is given by a PBC curve together with all the points enclosed by the PBC curve with a non-zero winding number\footnote{In this proof, we focus only on the continuous spectrum and ignore isolated points. In physics, such points are nothing but edge modes such as the topological zero modes. For Toeplitz matrices, which correspond to cases without internal degrees of freedom in the unit cell, the statement here does not suffer from this subtlety.} [Figure \ref{fig2}(a)].
Here, the winding number around a point $E\in\mathbb{C}$ is defined as
\begin{align}
W \left( E \right)
:= \int_{0}^{2\pi} \frac{dk}{2\pi i} \frac{d}{dk} \log \det \left( H \left( k \right) - E \right),\label{windingnumber}
\end{align}
where $H(k)$ is the Bloch Hamiltonian of $H$ under the PBC.
Corresponding to a point $E$ inside the PBC curve, the right (left) eigenstate is given by an exponentially localized boundary state for a negative (positive) winding number.
In step (ii), we begin with the following inclusion:
\begin{align}
    \sigma_{\mathrm{OBC}}(H)\subset\sigma_{\mathrm{SIBC}}(H),
\end{align}
where $\sigma$ denotes the spectrum.
Besides an exact justification, it is intuitively natural because the OBC is given by a combination of the SIBC and an additional boundary condition at the other boundary.
Once we admit this inclusion, we obtain another inclusion by applying the imaginary gauge transformation (\ref{imggauge}), which is a kind of similarity transformation, to the Hamiltonian.
The crucial point is that a similarity transformation changes the SIBC spectrum due to the infinite-dimensional nature of the matrix\footnote{Under the imaginary gauge transformation, the boundary of the SIBC spectrum is changed because the corresponding PBC Hamiltonian is also changed.
The change of the PBC curve can be treated by introducing a Bloch Hamiltonian for complex momenta, $H \left( k-i\log r \right)$ \cite{Lee-Thomale-19}.} [Figure \ref{fig2}(a)], while it does not change the OBC spectrum defined by the infinite-volume limit of the spectrum of a finite-dimensional matrix.
Thus, the original OBC spectrum is included in the transformed SIBC spectrum.
Since one can consider such an inclusion for an arbitrary imaginary gauge transformation $V_r$, we obtain
\begin{align}
\sigma_{\rm OBC} \left( H \right) \subset \bigcap_{r \in \left( 0, \infty \right)} \sigma_{\rm SIBC} \left( V^{-1}_{r} H V_{r} \right).\label{intersection_inclusion}
\end{align}
Now we are in a position to prove the theorem.
If the PBC spectrum has a non-trivial winding, one can find a SIBC mode with energy $E\in\mathbb{C}$ just inside the PBC curve, which is exponentially localized at the boundary.
This boundary mode is mapped to a plane wave via the imaginary gauge transformation $V_r$ with an appropriate $r$, and $E$ is on the edge of $\sigma_{\rm SIBC} \left( V^{-1}_{r} H V_{r} \right)$, or equivalently, the PBC spectrum of $V^{-1}_{r} H V_{r}$. 
Thus, the intersection $\sigma_{\rm SIBC}(H)\cap\sigma_{\rm SIBC} \left( V^{-1}_{r} H V_{r} \right)$ is smaller than the original SIBC spectrum $\sigma_{\rm SIBC}(H)$.
This procedure can be repeated unless the right-hand side of the inclusion $(\ref{intersection_inclusion})$ becomes a curve without winding, which implies that the OBC spectrum cannot have a nonzero winding [Figure \ref{fig2} (a)].

\subsection{The bulk-boundary correspondence\label{BBC-and-line-gap}}
The above consideration also clarifies how the conventional bulk-boundary correspondence breaks in the presence of the non-Hertmitian skin effect.
To see this,
let us consider a non-Hermitian lattice system with the bulk band spectrum forming spectral islands in the complex energy plane.
When the spectral islands are separated from each other, we say that the system has {\it line gaps} \cite{KSUS-19} (see Sect. \ref{classification}), and each island may support a topological number \cite{Esaki-11,Shen-18,KSUS-19,lein2020choosing}. 
Then, the bulk-boundary correspondence holds if one considers spectral islands in the OBC: If a spectral island in the OBC has a non-zero topological number, we have a corresponding topological boundary state.  

In the presence of the non-Hermitian skin effect, the spectral islands in the OBC are different from those in the PBC.
Therefore, the conventional bulk-boundary correspondence, which relates a topological number in the PBC to a boundary state, does not always hold. 
However, the breakdown has limitations.
By the argument in the previous subsection, the bulk OBC spectrum is always inside the PBC one, and thus
each bulk spectral island in the OBC is also inside a spectral island in the PBC. 
Thus, if the system has a line gap in the PBC, it also has a line gap in the OBC.
Since spectral islands in the PBC can be smoothly deformed into those in the OBC without closing line gaps,  
this means that 
the topological numbers of the spectral islands coincide between the PBC and the OBC under the same situation.
Therefore, once one has a well-defined topological number in the PBC, which is always true in the situation above, the corresponding topological number in the OBC takes the same value, and thus the conventional bulk boundary correspondence holds. 
Note that the PBC islands should merge before the gap closing (namely, a collision of spectral islands) in the OBC,
and thus the topological number in the PBC becomes ill-defined near a topological phase transition in the OBC.
The breakdown of the conventional bulk-boundary correspondence occurs in this situation.

\subsection{Correspondence between skin mode and topological exact zero mode}
Thus far, we have related the conventional skin effect with a non-trivial winding number of the PBC spectrum.
Actually, the same winding number characterizes the topological boundary zero modes of a Hermitian topological insulator.
The key idea is to introduce the doubled Hermitian Hamiltonian, which is also a tool for topological classifications of Floquet Hamiltonians \cite{roy-floquet-classification} and non-Hermitian Hamiltonians without boundaries \cite{Gong-18,KSUS-19}:
\begin{align}
\tilde{H}_E := 
\begin{pmatrix}
0&H-E\\
H^{\dagger}-E^*&0
\end{pmatrix},\label{doubleham}
\end{align}
where $E\in\mathbb{C}$ is a reference point.
Owing to the artificial doubling process, an additional chiral symmetry is imposed on the doubled Hermitian Hamiltonian as a constraint:
\begin{align}
    \Gamma \tilde{H}_E \Gamma^{-1}=-\tilde{H}_E
    ~~{\rm with}~~\Gamma=
    \begin{pmatrix}
    1&0\\
    0&-1
    \end{pmatrix}.
\end{align}
In the Altland-Zirnbauer classification \cite{Altland}, the symmetry class with one chiral symmetry is called class AIII, and the topological classification of insulators (gapped\footnote{The gapped nature of $\tilde{H}_E$ is defined by $\det \tilde{H}_E(k)\neq0$, or equivalently, $\det\left( H \left( k \right) - E \right)\neq0$.} phases) is given by $\mathbb{Z}$.
Actually, the integer topological invariant is nothing but the winding number of the PBC spectral curve of $H$ around the reference point $E$, given in Eq. (\ref{windingnumber}). 
Thus, the same winding number characterizes both a class-AIII topological insulator and a conventional skin effect.
In other words, the conventional skin effect gives one physical interpretation of a non-Hermitian topological classification \cite{Gong-18,Kawabata-19} in a special case (see Sect. \ref{classification} for details).
Remarkably, in this correspondence, one can relate the Hatano-Nelson model, which is the simplest model of the conventional skin effect, to the Su-Schrieffer-Heeger model \cite{SSH-79}, which is the simplest model of one-dimensional class-AIII topological insulators.

Moreover, the skin mode is not unrelated to the bulk-boundary correspondence. To see this, we begin with the SIBC.
If $E\in\sigma_{\rm SIBC}(H)\setminus\sigma_{\rm PBC}(H)$ and $W(E)<0$ for the PBC curve, one can find a boundary-localized right eigenstate $|E\rangle$ of $H$.
By using $|E\rangle$, one can construct a boundary zero mode of $\tilde{H}_E$ with negative chirality:
\begin{align}
    \tilde{H}_E
    \begin{pmatrix}
    0\\
    |E\rangle
\end{pmatrix}=0,~~~
\Gamma
    \begin{pmatrix}
    0\\
    |E\rangle
    \end{pmatrix}=-
    \begin{pmatrix}
    0\\
    |E\rangle
    \end{pmatrix}.\label{right-zero-mode}
\end{align}
This is nothing but a topological boundary zero mode of a class-AIII topological insulator.
Similarly, for a positive winding number, one can assign a topological boundary zero mode with positive chirality
\begin{align}
    \tilde{H}_E
    \begin{pmatrix}
    |E\rangle\!\rangle\\
    0
\end{pmatrix}=0,~~~
\Gamma
    \begin{pmatrix}
    |E\rangle\!\rangle\\
    0
    \end{pmatrix}=
    \begin{pmatrix}
    |E\rangle\!\rangle\\
    0
    \end{pmatrix},\label{left-zero-mode}
\end{align}
where $|\cdot\rangle\!\rangle$ denotes the Hermitian conjugate of the left eigenstate [see Eq. (\ref{rightleft})].

Basically, the same thing holds for the non-Hermitian skin modes under the OBC,
but there are several significant differences from the SIBC case.
One thing is that both the right and left eigenstates localized at the opposite boundaries can be defined at the same time, owing to the additional boundary. 
Correspondingly, the topological zero modes with opposite chirality appear at the opposite boundary of the doubled Hermitian system.
The more crucial difference is a subtle mismatch in the correspondence.
While $\tilde{H}_E$ corresponds to a topological insulator for any $E$ in a two-dimensional region on the complex plane $\sigma_{\rm SIBC}(H)\setminus\sigma_{\rm PBC}(H)$, $\sigma_{\rm OBC} \left( H \right)$ is given by a one-dimensional complex curve. Under the OBC with finite $L$, two topological boundary modes with opposite chirality, localized at the opposite sides, usually have a finite overlap, and the true eigenstates are their superpositions with small but finite energy that becomes zero only for $L\rightarrow \infty$.
Such topological modes are called quasi-zero modes, and the correspondence between a Hermitian eigenstate and a non-Hermitian eigenstate does not exactly hold.
In contrast, the topological boundary modes with exact zero energy have correspondence with the skin modes [Figure \ref{fig2}(b)], as in the case of the SIBC. This is the origin of the subtle mismatch mentioned above.

\subsection{Correspondence between pseudospectrum and topological quasi zero mode \label{pseudospectrum}}
The remaining question is what the non-Hermitian counterpart of the topological quasi-zero energy of $\tilde{H}_E$ in the correspondence is.
The answer is $\epsilon$-pseudospectrum $\sigma_{\epsilon}(H)$ \cite{Okuma-Sato-20}[Figure \ref{fig2}(b)], which is defined by the set of spectra of perturbed matrices\footnote{In numerical calculations, another equivalent definition \cite{Trefethen} $\sigma_{\epsilon}(H)=\{z\in\mathbb{C}~|~|\!|(z-H)^{-1}|\!|>\epsilon^{-1}\}$ is useful.} $H+\eta$ with $|\!|\eta|\!|<\epsilon$, where $|\!|\cdot|\!|$ is the matrix 2-norm or the largest singular value \cite{Trefethen}. 
While the $\epsilon$-pseudospectrum of a normal matrix $([H,H^{\dagger}]=0)$ such as a Hermitian matrix is just given by the $\epsilon$-neighborhood of the exact spectrum, that of a non-normal matrix $([H,H^{\dagger}]\neq0)$ can be much larger than the $\epsilon$-neighborhood.
In the present case, the following holds\footnote{The opposite limit corresponds to the OBC spectrum: $ \lim_{L\rightarrow\infty}\lim_{\epsilon\rightarrow0}\sigma_{\epsilon}(H^{(L)}_{\mathrm{OBC}})=\sigma_{\rm OBC}(H)$.} \cite{Trefethen, Gong-18,Okuma-Sato-20}:
\begin{align}
    \lim_{\epsilon\rightarrow0}\lim_{L\rightarrow\infty}\sigma_{\epsilon}(H^{(L)}_{\mathrm{OBC}})=\sigma_{\rm SIBC}(H),
\end{align}
where $H^{(L)}_{\mathrm{OBC}}$ is the size-$L$ OBC Hamiltonian.
Roughly speaking, the SIBC eigenstates approximate the states that correspond to the pseudospectrum, mentioned as $quasieigenstate$\footnote{Related to this terminology, one can also describe the $\epsilon$-pseudospectrum by another equivalent definition $\sigma_{\epsilon}(H)=\{z\in\mathbb{C}~|~|\!|(z-H)\bm{v}|\!|<\epsilon~\mathrm{for~some~unit~vector~}\bm{v}\}$, where $|\!|\cdot|\!|$ is the vector norm.} in Ref. \cite{Gong-18}.
Reference \cite{Okuma-Sato-20} showed the correspondence between topological quasi-zero modes with $\mathcal{O}(\epsilon)$ energy and $\epsilon$-pseudospectrum.
Thus, the pseudospectrum with infinitesimally small $\epsilon$ compensates for the subtle mismatch between the two-dimensional region with non-trivial topology and the OBC spectral curve in the complex plane. 
Note that the origin of the drastic difference between the OBC spectrum and pseudospectrum is the $nonlocal$ perturbations that connect the ends of the open chain.
In other words, the OBC spectrum is unstable against such nonlocal perturbations \footnote{Related to this mathematics, it is known that a numerical diagonalization of the non-normal matrices is sensitive to a rounding error because it can behave as a nonlocal perturbation \cite{Trefethen}.}, while it is robust against local perturbations.
By using the correspondence between Hermitian and non-Hermitian topology, one can relate this behavior to the following fact: The topological zero modes of a topological insulator are easily gapped out by connecting edges, while they are robust against the symmetry-preserving local perturbations.

\subsection{The Bauer-Fike theorem and the skin effect}
For a nonzero $\epsilon$ and a finite $L$, we have a useful bound for perturbations called the Bauer-Fike theorem \cite{Bauer-Fike}. 
When $H$ is diagonalizable, the theorem implies \cite{Trefethen} 
\begin{align}
%\sigma(H)+\Delta_\epsilon\subseteq
\sigma_\epsilon(H)\subseteq \sigma(H)+\Delta_{\epsilon\kappa(V)}, \label{eq:Bauer-Fike}   
\end{align}
with $\Delta_\delta=\{z\in \mathbb{C}:|z|<\delta\}$ and $\kappa(V)=\| V\|\cdot\|V^{-1}\|\ge 1$, where $V$ is a matrix diagonalizing $H$, $V^{-1}HV={\rm diag}(E_1,E_2,\dots)$. 
The condition number $\kappa(V)$ measures the non-normality of $H$ and give a bound for perturbed spectrum: If $H$ is normal so $V$ is unitary, it takes the minimal value $\kappa(V)=1$, so the perturbed spectrum stays the $\epsilon$-neighborhood of the exact spectrum.  
On the other hand, if $H$ is non-normal, it allows a larger perturbation of the spectrum. 
With biorthgonal right and left eigenstates $|E_n\rangle$, $|E_n\rangle\!\rangle$ of $H$, we have $V=(|E_1\rangle, |E_2\rangle, \dots)$ and $V^{-1}=(\langle\!\langle E_1|, \langle\!\langle E_2|,\dots)^t$. Because the matrix 2-norm satisfies $\|A\|\ge \sqrt{{\rm tr}(A^{\dagger} A)}/\sqrt{L}$ for a $L\times L$ matrix $A$ \cite{Golub-Van_Loan}, $\kappa(V)$ has a lower bound as 
\begin{align}
 \kappa(V)\ge 
\sqrt{\sum_i \langle E_i |E_i\rangle
\sum_j \langle\!\langle E_i |E_i\rangle\!\rangle}/L
=\sqrt{\sum_i \langle E_i |E_i\rangle
\sum_j \langle\!\langle E_i |E_i\rangle\!\rangle}
/\sum_k \langle\!\langle E_k|E_k\rangle .  
\end{align}
Therefore, $\kappa(V)$ can be huge if the skin effect occurs, where the right eigenstate $|E_i\rangle$ and left one $|E_i\rangle\!\rangle$ are localized on an opposite boundary. The extreme sensitivity of the spectrum against perturbations, which is suggested the bound $\epsilon \kappa(V)$ in Eq. (\ref{eq:Bauer-Fike}), opens a possible application of the skin effect to highly accurate sensors \cite{Budich-20}.

For a larger $\epsilon$, 
a variant of the Bauer-Fike theorem \cite{Trefethen}
\begin{align}
\sigma_{\epsilon}(H)\subseteq 
\sigma(H)+\Delta_{\epsilon+{\rm dep}(H)}   
\end{align}
with ${\rm dep}(H)=\sqrt{{\rm tr}(H^{\dagger}H)-\sum_i|E_i|^2}$
gives a more severe bound for perturbations.
The quantity ${\rm dep}(H)$ is called departure of normality \cite{Henrici}, and also measures the non-normality of $H$.
The skin effect results in a large difference between 
${\rm dep}(H)$ in the OBC and that in the PBC, and thus the difference also characterizes the skin effect \cite{Nakai-Okuma-Sato}.

We also note that the pseudospectrum plays an important role in non-Hermitian dynamics. This point will be discussed in Sect. \ref{dynamical}.

\section{Symmetry-protected skin effects under time-reversal symmetry}
So far we have shown that the PBC curve with a non-trivial winding number indicates the non-Hermitian skin effect.
Then a natural question arises: {\it Are all the non-Hermitian skin effects characterized by non-zero winding numbers?}
The answer is NO.

In modern physics, the quantum Hall effect is regarded as an example of broader concepts: topological insulator \cite{Kane-review,Zhang-review} or symmetry-protected topological phase \cite{senthil2015symmetry}.
Similarly, the conventional skin effect can be regarded as an example of a broader concept: symmetry-protected skin effects in general dimensions.
In this section, we consider one- and two-dimensional skin effects protected by a non-Hermitian time-reversal symmetry.
Instead of the $\mathbb{Z}$ winding number, they are characterized by $\mathbb{Z}_2$ topological invariant.

\subsection{Time-reversal symmetry in non-Hermitian systems \label{Def-of-TRS}}
For non-Hermitian Hamiltonian matrices, there are more types of symmetries than for Hermitian ones \cite{Bernard-LeClair-02,KSUS-19}. 
Among them, we focus on extensions of the time-reversal symmetry (TRS).
For a Hermitian Hamiltonian $H$, the TRS is defined as an antiunitary symmetry:
\begin{align}
    TH^*T^{-1}=H,\label{conjugateTRS}
\end{align}
where $T$ is taken as a unitary matrix.
Corresponding to the integer/half-integer spin of a particle, $TT^*=\pm1$ are assigned.   
One natural extension to non-Hermitian cases is the complex-conjugate-type TRS defined by the same equation (\ref{conjugateTRS}).
Since the transpose is not equivalent to the complex conjugation under non-Hermiticity, one can also consider another extension, namely, the transpose-type TRS \cite{Bernard-LeClair-02,KSUS-19}:
\begin{align}
    TH^TT^{-1}=H.\label{transposeTRS}
\end{align}
Depending on the situation, both TRSs may be physically relevant.
For example, the transpose-type TRS appears as a natural TRS of non-Hermitian Hamiltonians defined by the one-particle Green's function \cite{Yoshida-dagger-symmetry} or that of the quadratic Lindbladian \cite{Lieu-19}.

\subsection{$\mathbb{Z}_2$ skin effect in one dimension}
\begin{figure}
\begin{center}
 \includegraphics[width=12cm,angle=0,clip]{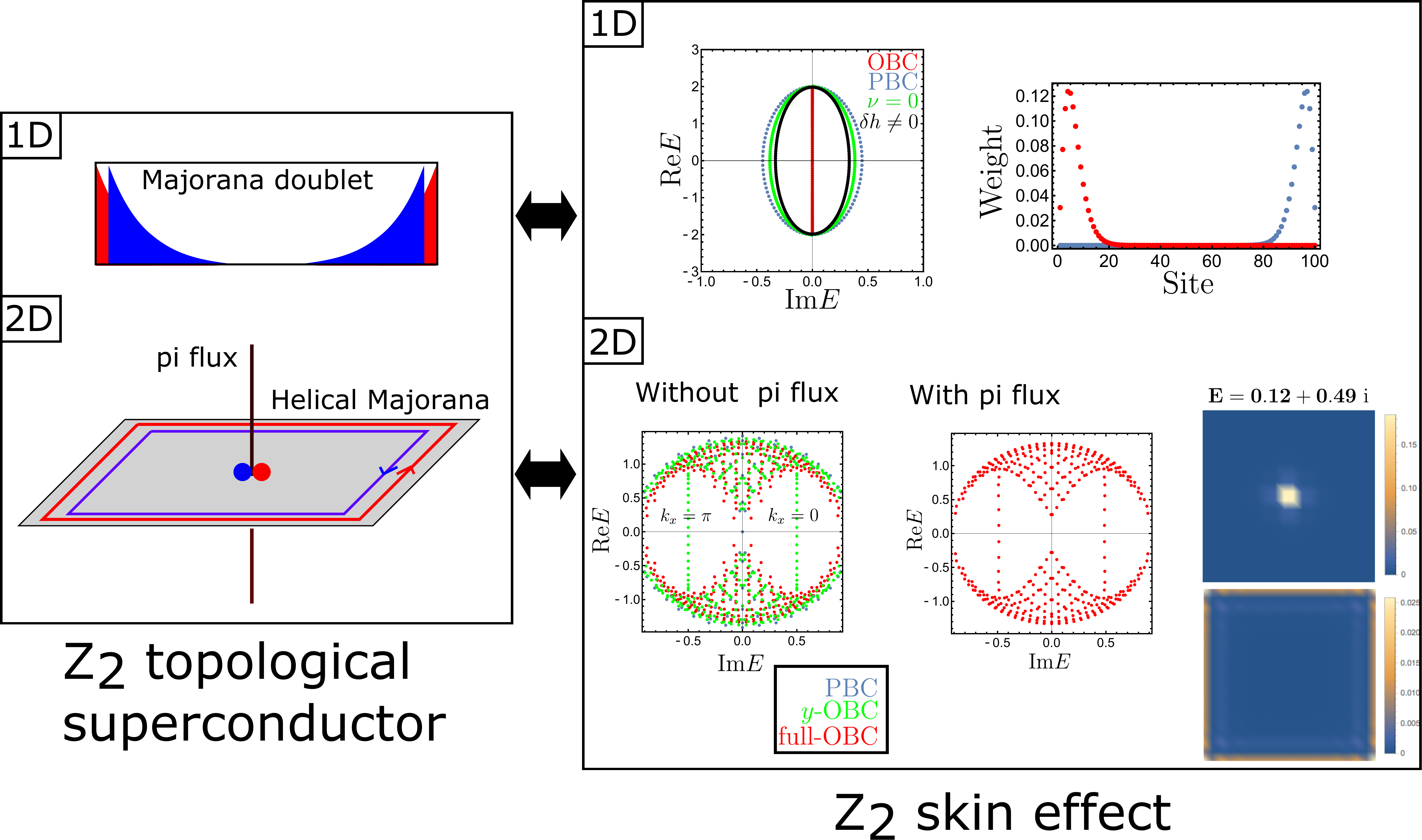}
 \caption{Relationship between $\mathbb{Z}_2$ topological superconductor and $\mathbb{Z}_2$ skin effects in one and two dimensions. The spectra and weight functions are adopted from Ref. \cite{OKSS-20}.}
 \label{fig3}
\end{center}
\end{figure}

In the previous section, we have introduced the doubled Hermitian Hamiltonian $\tilde{H}_E$ [see Eq.(\ref{doubleham})] to show the correspondence between a conventional skin effect and a class-AIII topological insulator.
In the presence of the transpose-type TRS, one can also define $\tilde{H}_E$ for arbitrary $E\in\mathbb{C}$ because the addition of terms proportional to the identity matrix does not break the relation (\ref{transposeTRS}).
For the cases with $TT^*=-1$, $\tilde{H}_E$ has a Hermitian TRS, an artificial chiral symmetry (CS), and their combination, namely, the particle-hole symmetry (PHS): 
\begin{align}
    {\rm TRS}&:~\tilde{T}\tilde{H}_E^*\tilde{T}^{-1}=\tilde{H}_E,~\tilde{T}\tilde{T}^*=-1,~\tilde{T}:=
    \begin{pmatrix}
    0&T\\
    T&0
    \end{pmatrix},
    \notag\\
    {\rm PHS}&:~\tilde{C}\tilde{H}_E^*\tilde{C}^{-1}=-\tilde{H}_E,~\tilde{C}\tilde{C}^*=1,~\tilde{C}:=
    \begin{pmatrix}
    0&-T\\
    T&0
    \end{pmatrix},\notag\\
    {\rm CS}&:~\tilde{\Gamma}\tilde{H}_E\tilde{\Gamma}^{-1}=-\tilde{H}_E,~\tilde{\Gamma}^2=1,~\tilde{\Gamma}:=
    \begin{pmatrix}
    1&0\\
    0&-1
    \end{pmatrix}.\label{trsdouble}
\end{align}
In the Altland-Zirnbauer classification \cite{Altland}, the set of these symmetries corresponds to the symmetry class DIII.
In one dimension, a class-DIII gapped Hermitian Hamiltonian (or the original non-Hermitian Hamiltonian) is classified by a $\mathbb{Z}_2$ topological invariant $\nu(E)\in\{0,1\}$ \cite{KSUS-19}
\begin{align}
\left( -1 \right)^{\nu \left( E \right)} := \mathrm{sgn} \left\{
\frac{ \mathrm{Pf} \left[ \left( H \left( \pi \right) - E \right) T \right] }{ \mathrm{Pf} \left[ \left( H \left( 0 \right) - E \right) T \right] }
\times \exp \left[ 
-\frac{1}{2} \int_{k=0}^{k=\pi} d \log \det \left[ \left( H \left( k \right) - E \right) T \right]
\right] 
\right\},\label{z2inv}
\end{align}
where $H(k)$ is again the Bloch Hamiltonian for the non-Hermitian Hamiltonian.
Under the OBC, the non-trivial phase ($\nu=1$) describes a topological superconductor with a Kramers doublet of Majorana fermions \cite{teo-kane}.
If the energy of the OBC boundary modes are exactly zero, one can relate the Majorana boundary states to eigenstates of $H$ as in the case of the conventional skin effect [Figure \ref{fig3}]:
\begin{align}
    \begin{pmatrix}
    0\\
    |E\rangle
\end{pmatrix},~
    \begin{pmatrix}
    T|E\rangle^*\\
    0
\end{pmatrix},~
    \begin{pmatrix}
    0\\
    T|E\rangle\!\rangle^*
\end{pmatrix},~
 \begin{pmatrix}
    |E\rangle\!\rangle\\
    0
\end{pmatrix}.\label{exactzeromodes}
\end{align}
The former/latter two modes form a Majorana doublet. Since the former and the latter pairs are localized at the opposite side, $|E\rangle~( T|E\rangle\!\rangle^*)$ and $|E\rangle\!\rangle~(T|E\rangle^*)$ are the right and left eigenstates of $H$ localized at the different side.
Thus, the right eigenstates $|E\rangle$ and $T|E\rangle\!\rangle^*$ satisfy the definition of skin modes and are localized at the different sides. 
In terms of the transpose-type TRS, they are in the relationship of the non-Hermitian Kramers pair \cite{Sato-11,KSUS-19}.
To emphasize the difference from the conventional skin effect, we call
this localization phenomenon the $\mathbb{Z}_2$ skin effect.
Note that the conventional winding number under the PBC cannot have a non-trivial winding number in the presence of the transpose-type TRS with $TT^*=-1$ because of the Kramers degeneracy.

In the following, we construct an explicit model of the $\mathbb{Z}_2$ skin effect.
Let us mimic the first example of the $\mathbb{Z}_2$ topological insulator protected by the TRS, called the Kane-Mele model \cite{Kane-Mele-05-QSH}.
This model was constructed by a stack of quantum anomalous Hall insulators with opposite Chern numbers.
In the absence of spin-orbit interactions, the Kane-Mele model is block diagonalized into spin sectors and is characterized by the integer spin Chern number. In the presence of the spin-orbit interaction that couples the spin sectors, the non-trivial phase is characterized by a $\mathbb{Z}_2$ topological invariant instead of the ill-defined spin Chern number.
Similarly, we begin with the stack of the Hatano-Nelson model (\ref{HNmodel}) with opposite winding numbers $W=\pm1$ \cite{Okuma-19}:
\begin{align}
H_{\mathbb{Z}} \left( k \right)
= \left( \begin{array}{@{\,}cc@{\,}} 
	H_{\rm HN} \left( k \right) & 0 \\
	0 &  H_{\rm HN}  \left( -k \right) \\ 
	\end{array} \right) 
= 2 t \cos k ~\sigma_0 + 2i g \sin k ~\sigma_{z},\label{spinhatano}
\end{align}
where $\sigma$'s represent the Pauli matrix of spin. Instead of the full Hamiltonian, we have specified the model by the Bloch Hamiltonian.
In this basis, the transpose-type TRS with $TT^*=-1$ for a Bloch Hamiltonian $H(k)$ is given by
\begin{align}
    TH^T(-k)T^{-1}=H(k),\label{sytrs}
\end{align}
where $T=i\sigma_y$.
Thus, $H_{\mathbb{Z}}(k)$ hosts the transpose-type TRS, and the total winding number (\ref{windingnumber}) cannot be non-trivial.
However, the block Hamiltonians in the spin sectors are given by the Hatano-Nelson models with opposite asymmetric hopping terms, which show the skin effects characterized by the opposite winding numbers for $tg\neq0$. This is an example of the skin effect protected by transpose-type symmetry.
Instead of the total winding number, the integer ``spin" winding number characterizes this skin effect. 

Next, we add a spin-orbit interaction $\Delta\geq0$ that connects the spin sectors \cite{OKSS-20}:
\begin{align}
H_{\mathbb{Z}_2} \left( k \right)
= \left( \begin{array}{@{\,}cc@{\,}} 
	H_{\rm HN} \left( k \right) & 2\Delta \sin k \\
	2\Delta \sin k &  H_{\rm HN}  \left( -k \right) \\ 
	\end{array} \right)
= 2 t \cos k + 2 \Delta \left( \sin k \right) \sigma_{x} + 2i g \left( \sin k \right) \sigma_{z}.
\end{align}
Since the spin-rotational symmetry is completely broken, the winding number in each spin sector cannot be defined for non-zero $\Delta$.
Even for this case, we can still define the $\mathbb{Z}_2$ invariant (\ref{z2inv}) due to the presence of TRS (\ref{sytrs}).
The PBC spectrum is given by two bands $E_{\pm} \left( k \right) = 2t \cos k \pm 2i \sqrt{g^2-\Delta^2} \sin k$, both of which describe an ellipse in the complex plane for $|g|>\Delta$ and $t\neq0$ \cite{OKSS-20}.
Correspondingly, the $\mathbb{Z}_2$ invariant (\ref{z2inv}) is not trivial for this parameter region, which should lead to the non-Hermitian skin effect.
The numerical diagonalization for a finite size ($L=100$) shows the presence of skin effect with Kramers pairs localized at the opposite sides [Figure \ref{fig3}]. 
This is an example of symmetry-protected skin effects characterized by a $\mathbb{Z}_2$ topological invariant (\ref{z2inv}).

We here give several remarks about the symmetry-protected skin effects.
First, the symmetry-protected nature appears as an instability of the skin effect against symmetry-breaking perturbations [$\delta h\neq0$ in Figure \ref{fig3}].
In the infinite-volume limit, this instability is understood by the failure of the standard non-Bloch band theory under some symmetry protection \cite{Okuma-19}.
In the presence of the symmetry that block-diagonalizes $H(k)$ such as Eq. (\ref{spinhatano}), the corresponding OBC spectrum is determined not by the non-Bloch band theory for the total Hamiltonian but by that for each block.
Under symmetry-breaking terms that connect the blocks, however, the OBC spectrum is determined by the non-Bloch band theory for the total Hamiltonian.
Remarkably, the latter spectrum for an infinitesimally small perturbation can be different from the former spectrum.
For example, the OBC spectrum of Eq. (\ref{spinhatano}) under an infinitesimally small transverse magnetic field becomes identical to the PBC one \cite{Okuma-19}.
In the case of the transpose-type TRS without spin-rotational symmetry, a similar discussion can be applied by introducing a modified non-Bloch band theory with Kramers doublet \cite{KOS-20}. 
In the language of doubled Hermitian Hamiltonian, this phenomenon is understood by the fact that the exact zero modes acquire the finite gap under a symmetry-breaking term, which destroys the correspondence between the skin modes and the exact zero modes mentioned above \footnote{In a finite system, the threshold of the perturbation strength for the instability is exponentially small with respect to the system size.
In terms of the correspondence between the Hermitian and non-Hermitian topology, this threshold corresponds to the amount of perturbation that changes the energies of boundary modes significantly.
}.

Second, the $\mathbb{Z}_2$ nature is checked by a stack of two copies of the system with the skin effect whose topological number is unity. The skin effect in such a stack is fragile even against symmetry-preserving terms [$\nu=0$ in Figure \ref{fig3}].
This behavior can be understood by the fact that a skin mode of one Kramers doublet localized at the left side can be mixed with a skin mode of the other doublet localized at the right side.

These are the basic properties of symmetry-protected skin effects in one spatial dimension.
One can apply a similar consideration based on the doubled Hermitian Hamiltonian to higher dimensions.
However, more careful discussion about the boundary condition is needed to characterize a higher-dimensional non-Hermitian skin effect.

\subsection{$\mathbb{Z}_2$ skin effect in two dimensions}
We consider two-dimensional non-Hermitian Hamiltonians with the transpose-type TRS with $TT^*=-1$.
Again, a doubled Hermitian Hamiltonian $\tilde{H}_E$ describes class-DIII superconductors and is classified by a $\mathbb{Z}_2$ invariant under the PBC \cite{KSUS-19}:
\begin{equation} 
\begin{split}
( -1 )^{\nu \left( E \right)} :=& \prod_{\mathsf{X}=\mathrm{I, II}} 
\mathrm{sgn} 
\left\{
\frac{ \mathrm{Pf} 
\left[ 
\left( H 
\left( {\bm k}_{\mathsf{X}+} 
\right) - E 
\right) T 
\right] 
}
{ \mathrm{Pf} 
\left[ 
\left( H 
\left( {\bm k}_{\mathsf{X}-} 
\right) - E 
\right) T 
\right] 
}
\right.
\notag\\
&
\left.
\times \exp 
\left[ 
-\frac{1}{2} \int_{{\bm k} = {\bm k}_{\mathsf{X}-}}^{{\bm k} 
= {\bm k}_{\mathsf{X}+}} d \log \det 
\left[ 
\left( H \left( {\bm k} \right) - E \right) T 
\right]
\right] 
\right\},
\end{split} 
\label{2dz2inv}
\end{equation}
where $\bm{k}$ is a two-dimensional momentum, and (${\bm k}_{\mathrm{I}+}, {\bm k}_{\mathrm{I}-}$) and (${\bm k}_{\mathrm{II}+}, {\bm k}_{\mathrm{II}-}$) are two pairs of time-reversal-invariant momenta.
For the non-trivial phase $\nu \left( E \right)=1$, topological boundary modes are givien by helical Majorana edge modes \cite{teo-kane}.
If there exist exact zero modes,
they are given by the same expression as Eq. (\ref{exactzeromodes}).
Unlike one dimension, however, the exact zero modes cannot be found under the full OBC, i.e., the OBC in both the $x$ and $y$ directions.
This is because a topological gapless mode at the closed boundary of a higher-dimensional topological insulator/superconductor effectively feels the curvature of the boundary, and the ``zero" modes inevitably have finite energy in the order of $1/L$ \cite{Imura-12,Parente-11,DLee-09}.
Correspondingly, the $\mathbb{Z}_2$ skin effect does not occur for this boundary condition.
To observe a two-dimensional $\mathbb{Z}_2$ skin effect, one should choose boundary conditions that allow the exact topological boundary zero modes of $\tilde{H}_E$.
At least two boundary conditions for this purpose are known for higher-dimensional topological insulators/superconductors: (i) the OBC in one direction and the PBC in the other directions, and (ii) the full OBC with topological defects \cite{teo-kane}.

In the following, we describe the properties of the two-dimensional $\mathbb{Z}_2$ skin effect by using the massless Dirac Hamiltonian coupled to valley-dependent dissipation characterized by $\Gamma\in\mathbb{R}$ \cite{OKSS-20}:
\begin{align}
H \left( \bm{k} \right)=(\sin k_x \sigma_x+\sin k_y\sigma_y)+i\Gamma(\cos k_x+\cos k_y).\label{2dskineffect}
\end{align}
For non-zero $\Gamma$,
the set $\{E\in\mathbb{C}~|~\nu(E)=1\}$ occupies a two-dimensional region of the complex plane.
The complex spectrum of this system is shown in Figure \ref{fig3} for various boundary conditions \cite{OKSS-20}.
In case (i), the OBC is imposed in the $y$ direction. In case (ii), the topological defect is given by a $\pi$ flux (or a half flux quantum) at the central unit cell.
In contrast to the full PBC and full OBC, there exist skin modes for both cases.
Remarkably, only $\mathcal{O}(L)$ modes of total $\mathcal{O}(L^2)$ modes show the $\mathbb{Z}_2$ skin effect in which the spectrum has no winding, and they are separated from the other modes surrounding them. In case (i), $\mathcal{O}(L)$ skin modes are also interpreted as the one-dimensional $\mathbb{Z}_2$ skin modes of a one-dimensional Hamiltonian $H(k_x)$ at the time-reversal-symmetric points $k_x=0,\pi$, where $H(k_x)$ is the Fourier transform of $H$ in the $x$ direction.
The skin modes are localized at the boundaries in the $y$ direction.
In $D$-dimensional topological insulators/superconductors under the boundary condition (i), only $\mathcal{O}(1)$ modes of the total $\mathcal{O}(L^{d-1})$ modes of the surface Dirac Hamiltonian correspond to the exact zero modes.
Thus, $D$-dimensional skin effect has only $\mathcal{O}(L)$ skin modes\footnote{ The Hamiltonian (\ref{2dskineffect}) with $\Gamma=1$ has $\mathcal{O}(L)$-fold algebraic degeneracy at $E=\pm i$, which are exceptional points (see Sect.\ref{EP}). Three-dimensional cases were investigated in Ref. \cite{Terrier-Kunst-20}. }. In case (ii), $\mathcal{O}(L)$ skin modes are localized at the boundary and the topological defect.
In contrast to the other skin effects, this case shows a non-Hermitian localization in the radial direction [Figure \ref{fig3}].
Note that this two-dimensional localization purely originates from the non-Hermiticity, while the higher-order skin effect in Sect. \ref{hose} is reduced to the combination of Hermitian and non-Hermitian localizations.

\section{Physical interpretation of non-Hermitian topological classification\label{classification}}
We here review several topics related to non-Hermitian topological classification.
In particular, the anomaly interpretation provides another insight into symmetry-protected skin effects, in addition to the topological aspects discussed above. 
\subsection{Review of Non-Hermitian topological classification}
In Hermitian physics, a standard classification of non-interacting topological insulators/superconductors is given by the K-theoretical classification of a momentum-resolved Hamiltonian matrix $H(\bm{k})$ \cite{Schnyder-08,Kitaev-09,Ryu-10,Schnyder-Ryu-review}, where $\bm{k}$ is a momentum on a sphere or a torus.
In this field, the ten-fold Altland-Zirnbauer class \cite{Altland}, in which each class is specified by the combination of TRS, PHS, and CS, is regarded as the most fundamental symmetry class [for example, see Eq.(\ref{trsdouble})].
The gapped nature of insulators/superconductors is captured by the condition $\det H(\bm{k})\neq0$, which means the absence of zero energy in the bulk spectrum.
Gong et al. adopted the same condition for a non-Hermitian $H(\bm{k})$ as the definition of a non-Hermitian gapped phase \cite{Gong-18}.
This type of complex energy gap was named {\it point gap} in Ref. \cite{KSUS-19} [Figure \ref{fig4}(a)].
They showed that the topological classification of point-gapped phases is given by the K-theoretical classification of the corresponding doubled Hermitian Hamiltonian
\begin{align}
\tilde{H}(\bm{k}) := 
\begin{pmatrix}
0&H(\bm{k})\\
H^{\dagger}(\bm{k})&0
\end{pmatrix},
\end{align}
by noticing the equivalence between $\det H(\bm{k})\neq0$ and $\det \tilde{H}(\bm{k})\neq0$. 
After this study, Refs. \cite{KSUS-19,ZL-19} applied this classification scheme for 38-fold Bernard-LeClair class, which was originally introduced to describe non-Hermitian random matrices \cite{Bernard-LeClair-02}.
Reference \cite{KSUS-19} proposed another non-Hermitian gapped structure named the real/imaginary {\it line-gapped} phase [Figure \ref{fig4}(a)].
In this scheme, the gapped phase is defined as a spectral structure whose real/imaginary parts are nonzero. 
Furthermore, the concept of the line gap has been generalized to situations where several spectral islands exist in the complex energy plane \cite{lein2020choosing}.

Next, we consider physical interpretations of the non-Hermitian gapped phases.
In the case of the Hermitian classification, 
the non-trivial phase of the classification corresponds to a topological insulator/superconductor, in which the bulk non-trivial topology indicates the presence of topological gapless boundary modes. 
Actually, the real/imaginary line-gapped phases can be adiabatically connected to Hermitian/anti-Hermitian gapped Hamiltonians without breaking the fundamental symmetries and closing the line gap \cite{Esaki-11,KSUS-19,ashida-gong-20}.
Thus, the non-trivial topology in a line-gapped phase is essentially the same as that in a Hermitian gapped phase and indicates the bulk-boundary correspondence with a certain modification (see Sects. \ref{breakdown-of-bbc} and \ref{BBC-and-line-gap}).
In contrast, it is not easy to say something about the point-gapped phases because the non-trivial topology indicates the bulk-boundary correspondence of the doubled Hermitian Hamiltonian $\tilde{H}$, not of the original Hamiltonian $H$.
To the best of the authors' knowledge, there is no unified physical interpretation for the point-gapped phases\footnote{There are even cases where the point-gap topological invariant is reminiscent of a line-gap topological invariant, whose origin is essentially Hermitian topology \cite{OKSS-20}.}.
In the following, we give several interpretations for limited cases.

\subsection{Classification of non-Hermitian skin effects}
The symmetry-protected skin effects give one physical interpretation of the point-gap topological classification for classes without symmetry or with only transpose-type TRS ($TT^{*}=\pm1$) \cite{OKSS-20,Okuma-Sato-21}. 
In these three classes, also called class A, $\mathrm{AI}^{\dagger}$, and $\mathrm{AII}^{\dagger}$ \cite{KSUS-19}, the addition of arbitrary reference energy $E\in\mathbb{C}$ to $H$ does not break the given symmetry, allowing us to construct symmetry-protected skin effects discussed above \cite{Okuma-Sato-21}.
The classification of skin effects for these three classes is highlighted in red in Table \ref{table1}.
The above discussion does not mean that symmetry-protected skin effects occur only in these three classes, because the topological invariants of the three classes can be non-zero in other classes.

\subsection{Non-Hermitian topology and quantum anomaly}
\subsubsection{Anomaly interpretation for AZ$^{\dagger}$ symmetry}
As we noted in Sect. \ref{Def-of-TRS}, there are two types of non-Hermitian TRS, i.e., complex-conjugate-type and transpose-type TRS.
Similarly, one can define complex-conjugate-type and transpose-type PHS.
The ten-fold Altland-Zirnbauer (AZ) class is defined by the combination of complex-conjugate-type TRS and transpose-type PHS, while the ten-fold AZ$^{\dagger}$ class\footnote{In this paper, we call the combination of two-fold complex AZ class \cite{KSUS-19} and eight-fold real AZ$^{\dagger}$ class \cite{KSUS-19} the ten-fold AZ$^{\dagger}$ class.} is defined by the combination of transpose-type TRS and complex-conjugate-type PHS.

\begin{figure}
\begin{center}
 \includegraphics[width=12cm,angle=0,clip]{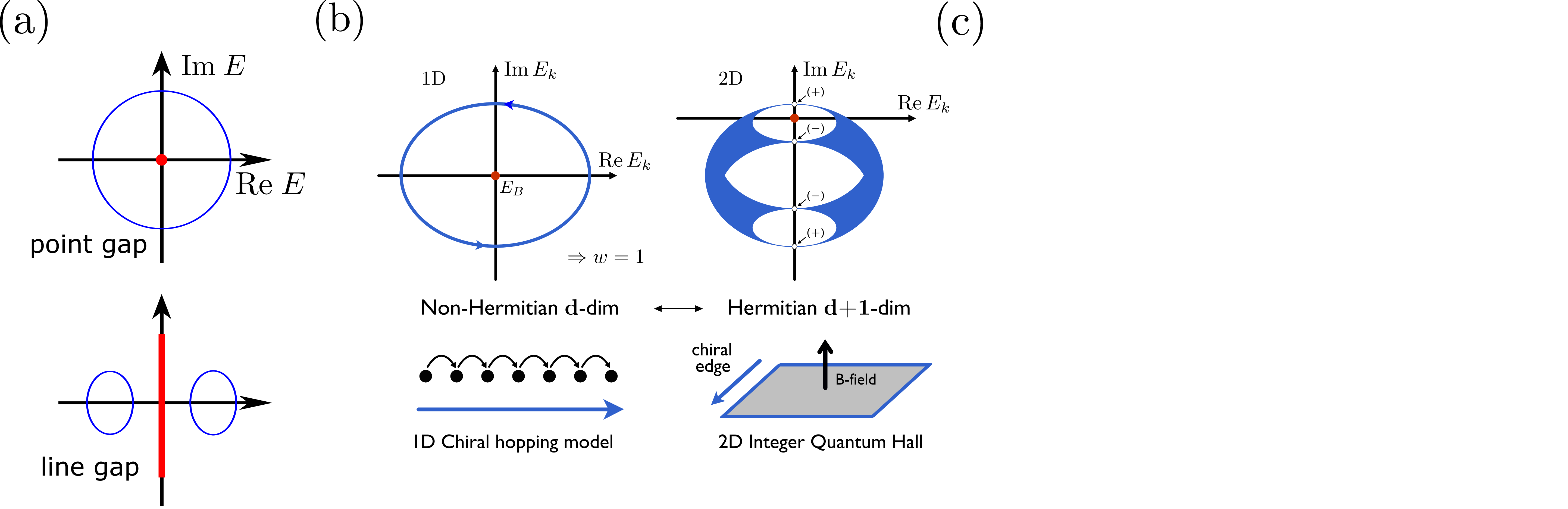}
 \caption{(a) Definitions of non-Hermitian gapped spectra (blue). A point/line-gapped spectrum does not contain red point/line.  (b) Correspondence between class-A point-gap topology and chiral edge mode at the boundary of quantum Hall effect, adopted from Ref. \cite{Lee-Vishwanath-19}. (c) Exceptional point (EP) and its characterization by point and line gap, adopted from Ref. \cite{KBS-19}.  }
 \label{fig4}
\end{center}
\end{figure}

As far as the AZ$^{\dagger}$ class is concerned, one can give a physical interpretation of the non-Hermitian topological classification, in the absence of boundaries.
Lee et al. pointed out that a point-gap topological invariant of the PBC curve in the class $s^{\dagger}\in$AZ$^{\dagger}$ counts the number of anomalous gapless modes whose imaginary parts are large enough \cite{Lee-Vishwanath-19}.
Here, the anomalous gapless modes in $D$ dimensions are the gapless modes that cannot appear on the bulk of a lattice due to the quantum anomaly but
can appear on a $D$-dimensional boundary of a $(D+1)$-dimensional Hermitian topological insulator/superconductor in the corresponding Hermitian class $s\in$AZ.
In linear dynamics described by a non-Hermitian Hamiltonian, the eigenvalues with large imaginary parts are relevant to the long-time dynamics. 
In this sense, the dynamics described by a point-gap non-trivial Hamiltonian are governed by the anomalous gapless modes in the long-time limit.
For example, the chiral modes, which appear on the edge of a quantum Hall insulator, describe the relevant modes in long-time dynamics of one-dimensional class-A non-trivial systems [Figure \ref{fig4}(b)].
More precise formulae are summarized as the extended Nielsen-Ninomiya theorem in Ref. \cite{Bessho-Sato-20}.

\begin{table}[]
\caption{Point-gap topological classification in AZ$^{\dagger}$ class \cite{KSUS-19}. The red-colored rows can also be regarded as the skin-effect classification. }
\label{table1}
\centering
$$
\begin{array}{c|ccc|ccccccccc}
\mbox{~AZ}^{\dagger}~&T&C&\Gamma&0&1&2&3&4&5&6&7\\
\hline \hline
\textcolor{red}{{\rm A}}&0&0&0&0&\textcolor{red}{\mathbb{Z}}&0&\textcolor{red}{\mathbb{Z}}&0&\textcolor{red}{\mathbb{Z}}&0&\textcolor{red}{\mathbb{Z}}\\ 
{\rm AIII}&0&0&1&\mathbb{Z}&0&\mathbb{Z}&0&\mathbb{Z}&0&\mathbb{Z}&0\\
\hline
\textcolor{red}{{\rm AI}^{\dagger}}&1&0&0&0&0&0&\textcolor{red}{2\mathbb{Z}}&0&\textcolor{red}{\mathbb{Z}_2}&\textcolor{red}{\mathbb{Z}_2}&\textcolor{red}{\mathbb{Z}}\\
{\rm BDI}^{\dagger}&1&1&1&\mathbb{Z}&0&0&0&2\mathbb{Z}&0&\mathbb{Z}_2&\mathbb{Z}_2\\
{\rm D}^{\dagger}&0&1&0&\mathbb{Z}_2&\mathbb{Z}&0&0&0&2\mathbb{Z}&0&\mathbb{Z}_2\\
{\rm DIII}^{\dagger}&-1&1&1&\mathbb{Z}_2&\mathbb{Z}_2&\mathbb{Z}&0&0&0&2\mathbb{Z}&0\\
\textcolor{red}{{\rm AII}^{\dagger}}&-1&0&0&0&\textcolor{red}{\mathbb{Z}_2}&\textcolor{red}{\mathbb{Z}_2}&\textcolor{red}{\mathbb{Z}}&0&0&0&\textcolor{red}{2\mathbb{Z}}\\
{\rm CII}^{\dagger}&-1&-1&1&2\mathbb{Z}&0&\mathbb{Z}_2&\mathbb{Z}_2&\mathbb{Z}&0&0&0\\
{\rm C}^{\dagger}&0&-1&0&0&2\mathbb{Z}&0&\mathbb{Z}_2&\mathbb{Z}_2&\mathbb{Z}&0&0\\
{\rm CI}^{\dagger}&1&-1&1&0&0&2\mathbb{Z}&0&\mathbb{Z}_2&\mathbb{Z}_2&\mathbb{Z}&0\\
\hline
\hline
\end{array}
$$
\end{table}

\subsubsection{Anomaly interpretation of non-Hermitian skin effects}
In classes A, AI$^{\dagger}$, and AII$^{\dagger}$, 
the symmetry-protected skin effects give the physical interpretation of non-trivial non-Hermitian topology under the OBC, while the anomalous gapless modes do under the PBC. 
These facts indicate the presence of the anomaly interpretation of the symmetry-protected skin effects.
In the anomaly interpretation, the skin effects are related to fermion production caused by the quantum anomaly \cite{Okuma-Sato-21}.
For example, a conventional skin effect is related to a charge accumulation at the boundary caused by a chiral current. 
The anomaly interpretation also gives an intuitive reason why higher-dimensional skin effects occur in the presence of the topological defect.
In quantum field theory, the combination of the quantum anomaly and the topological defect causes the fermion production at the topological defect.
The most famous example is the Rubakov-Callan effect (or monopole catalysis), which was originally introduced as the mechanism for proton decay in a SU(5) grand unified theory \cite{Rubakov,callan}.
The corresponding skin effect is realized in a class-A three-dimensional Weyl Hamiltonian with valley-dependent dissipation in the presence of a magnetic monopole \cite{Okuma-Sato-21}.
The chiral magnetic skin effect in Ref. \cite{Bessho-Sato-20} can be related to a typical fermion production mechanism called the chiral magnetic effect \cite{fukushima2008chiral}.

\subsection{Non-Hermitian topology and degeneracy points\label{EP}}
In the Hermitian topological classification, gapless structures in momentum space such as Dirac and Weyl point/line nodes are also of great interest \cite{stab-of-fermi-surface, matsuura2013protected,Zhao-PRL-13,Zhao-PRB-14,Zhao-PRL-16,Zhao-PRL-17,Morimoto-semimetal,Kobayashi-14,Shiozaki-Sato-14,CKC-14}.
In a $D$-dimensional parameter space, a robust $d$-dimensional symmetry-protected gapless structure is characterized by a topological gapped structure on a $(D-d)$-dimensional sphere surrounding the gapless structure.
On the basis of similar topological classification on sphere, Ref. \cite{KBS-19} gave point-gap and line-gap topological classifications around the exceptional points\footnote{Note that  non-trivial classifications do not always ensure the existence of exceptional points.}, at which the parameterized non-Hermitian Hamiltonian is not diagonalizable \cite{kato2013perturbation,berry2004physics,heiss2012physics} [Figure \ref{fig4}(c)]. 
According to Ref. \cite{KBS-19}, the point/line gap is open/closed around an exceptional point. This statement is based on the observation that exceptional points are connected via a line-gapless structure called the bulk Fermi arc \cite{Kozii-17,zhou-18,Papaj-19} [Figure \ref{fig4}(c)]. 
Such topological classifications describe symmetry-protected exceptional structures with various dimensionality such as exceptional points and rings \cite{Bergholtz-review,Kozii-17, zhou-18, Papaj-19, Zhou-19,Shen-18, Zyuzin-18, Yoshida-18, Yoshida-19, Bergholtz-19, Kimura-19, Zhen-15, Xu-17, Okugawa-19, Budich-19, Cerjan-19}.

\section{SKIN EFFECTS IN VARIOUS SITUATIONS}
\subsection{Skin effects without asymmetric hopping}
\begin{figure}
\begin{center}
 \includegraphics[width=12cm,angle=0,clip]{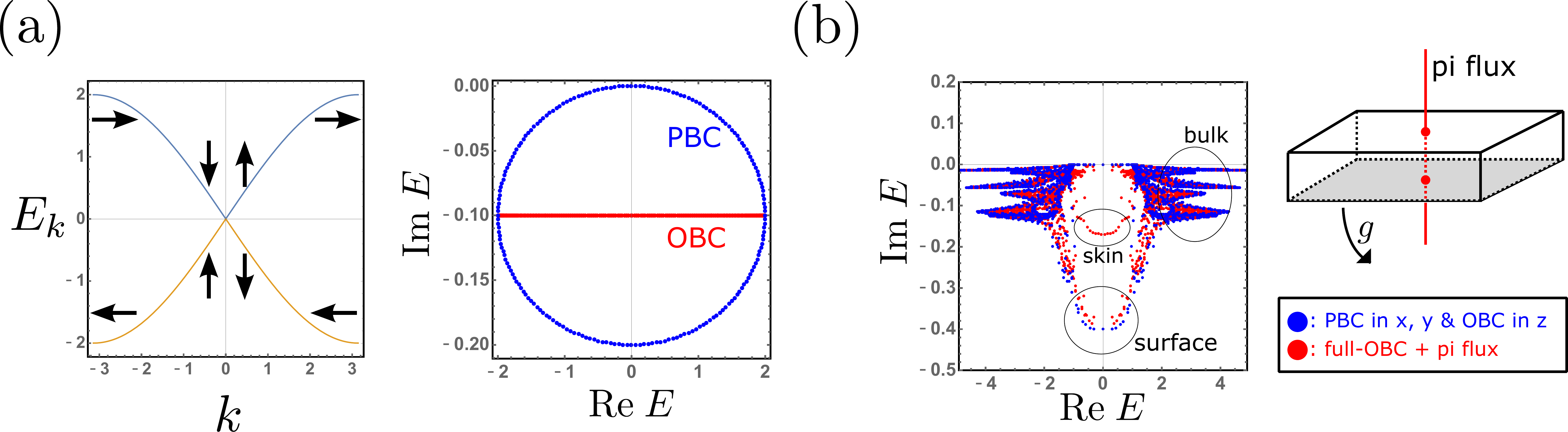}
 \caption{Skin effects without momentum-dependent dissipation. (a) One-dimensional skin effect in spin-momentum-locked bands of the Hamiltonian $H(k)=\sin k\sigma_z+(1-\cos k)\sigma_x$. A spin-dependent momentum-independent dissipation $i g (-1+\sigma_z)$ effectively behaves as momentum-dependent dissipation. The model parameters are $L=100$ and $g=0.1$. (b) Two-dimensional $\mathbb{Z}_2$ skin effect realized on the surface of a three-dimensional topological insulator $H(\bm{k})=\sum_{i=1}^{3} \sin k_i\gamma_i+(m+\sum_{i=1}^{3}\cos k_i)\gamma_0$, where $\gamma$s are the Gamma matrices satisfying $\{\gamma_\mu,\gamma_{\nu}\}=\delta_{\mu\nu}$. The constant dissipation $-i g$ is introduced at the bottom surface, and the pi flux goes through the center. The system size is $20\times20\times6$, $m=-2$, and $g=0.4.$}
 \label{fig5}
\end{center}
\end{figure}

Non-Hermitian skin effects are experimentally relevant in classical systems
\cite{Brandenbourger-19-skin-exp,Ghatak-19-skin-exp,Helbig-19-skin-exp,Hofmann-19-skin-exp}.
While the implementation of asymmetric hopping is not easy in quantum systems, this does not forbid the quantum implementation because non-Hermitian skin effects only require the spectral topology.
Namely, on-site dissipation can also induce non-Hermitian skin effects, as emphasized in Refs. \cite{Yi-Yang-20}.
In fact, the non-Hermitian skin effect in the discrete-time non-unitary quantum walk was experimentally realized by mode-selective loss \cite{Xiao-19-skin-exp}. 
For an intuitive understanding, let us consider a one-dimensional two-band Hermitian Hamiltonian $H(k)=\bm{d}(k)\cdot\bm{\sigma}$, where $\bm{d}\in\mathbb{R}^3$, and $\sigma$'s are the Pauli matrices that represent ``spin" degrees of freedom.
The dispersion is given by $E_{\pm}(k)=\pm |\bm{d}(k)|$, and the spin direction of each eigenstate is parallel/anti-parallel to $\bm{d}(k)$.
Under momentum-independent spin-dependent non-Hermiticity, each band effectively feels momentum-dependent dissipation \cite{Okuma-Sato-21}.
For $\bm{d}(k)$ that realizes the PBC curve with a non-trivial winding number, one can realize the conventional skin effect by the combination of the spin-momentum locked band structure and the on-site non-Hermiticity [Figure \ref{fig5}(a)]. 
This construction is analogous to an implementation of a topological superconductor in which the $p$-wave paring is effectively realized by the combination of the spin-momentum-locked band structure and the $s$-wave paring \cite{Sato-Ando,Sato-Fujimoto,Lutchyn,O-R-vO}.
One can also use the combination of topological boundary states and boundary-dependent dissipation as a source of non-Hermitian skin effects \cite{Okuma-Sato-21}.
For example, let us consider a thin film of a three-dimensional topological insulator.
The low-energy effective model of this system is given by two two-dimensional Dirac cones with opposite chiralities at the top and the bottom surfaces.
By setting a surface-dependent dissipation, one can realize the effective valley-dependent dissipation that leads to the two-dimensional $\mathbb{Z}_2$ skin effect [Figure \ref{fig5}(b)].

\subsection{Skin effects in higher dimensions\label{hose}}
In the previous sections, we have discussed $D$-dimensional skin effects whose origin is a $D$-dimensional topology.
In addition to such {\it intrinsic} $D$-dimensional skin effects, one can also consider the $D$-dimensional skin effect whose origin is a $(d<D)$-dimensional topology. For example, under the OBC in the $x$ direction and PBC in the other direction,
the Hamiltonian is block diagonalized into the Bloch Hamiltonian $H(\bm{k}_{\perp}\in\mathbb{R}^{D-1})$, which can be regarded as a one-dimensional system.
For each $\bm{k}_{\perp}$, one can define a conventional non-Hermitian skin effect whose origin is the spectral winding.
In this context, Ref. \cite{Hofmann-19-skin-exp}
theoretically and experimentally discussed a relationship between the emergence of exceptional points and a skin effect in a two-dimensional Dirac system.

Another remarkable direction is the corner skin modes at a higher-dimensional lattice, which are realized as the combination of Hermitian and non-Hermitian localizations.
Reference \cite{Lee-Li-Gong-19} proposed the hybrid higher-order skin-topological mode as the combination of Hermitian topological gapless state and the non-Hermitian skin effect, and Ref. \cite{zou2021observation} experimentally realized it on electric circuits.
References \cite{Okugawa-takahashi-yokomizo,KSS-20,Fu-Hu-Wan-21} investigated higher-order skin effects whose doubled Hermitian Hamiltonians are higher-order topological insulators\footnote{These skin modes are related to Hermitian edge states of semi-metals such as graphene. In this sense, Hermitian and non-Hermitian localizations coexist.} \cite{benalcazar2017quantized,benalcazar2017electric,langbehn2017reflection,song2017d}.
In addition to these studies, higher-order localizations have been studied in various contexts \footnote{Correspondingly, the terminology ``higher-order skin effects" are often used in various situations. }.
For example, the combinations of non-Hermitian skin effect and higher-order topological modes have been investigated both theoretically \cite{Liu-19,Edvardsson-19,GLSH-21,GLS-21} and experimentally \cite{wu2020observation,zhang2021observation}.

\section{NON-HERMITIAN DYNAMICS AND SPECTRAL TOPOLOGY\label{dynamical}}

Before closing, we relate the spectral topology of non-Hermitian tight-binding models to non-Hermitian dynamics.
Let us consider the non-Hermitian Schor\"{o}dinger equation with a non-Hermitian Hamiltonian $H$:
\begin{align}
    i\frac{\partial}{\partial t}|\psi(t)\rangle=H|\psi(t)\rangle,\label{nonhermitiansch}
\end{align}
where $|\psi(t)\rangle$ is a wavefunction whose norm depends on time.
In open quantum systems, this equation describes various situations, such as the Lindblad equation \cite{Lindblad} written in a quadratic form of field operators \cite{Prosen-2008,Prosen-2010} and the continuous Lindblad dynamics without quantum jumps \cite{Daley-14}.
In classical systems,
this equation has been investigated in various fields, including fluid mechanics \cite{Trefethen} and network science \cite{Asllani-2018}.
If $H$ is a normal matrix ($[H,H^{\dagger}]=0$),
the dynamics of the norm is always governed by the imaginary part of the spectrum.
If not ($[H,H^{\dagger}]\neq0$), however, the relaxation dynamics is more non-trivial.
Remarkably, the right eigenstates of a non-normal matrix do not span the whole Hilbert space.
If the initial state is in or close to the subspace spanned by the right eigenstates,
the non-normal dynamics is not different from the normal dynamics.
For more general initial states, however, there is a transient time $(0<t<\infty)$ for the state to have a large overlap with that subspace, and the transient dynamics is governed by the pseudospectrum discussed in Sect. \ref{pseudospectrum}, instead of the spectrum.
In the long-time limit $(t\rightarrow\infty)$, non-normal dynamics is governed by the spectrum, as in the case of a normal matrix.
Due to the pseudospectral nature, amplification in transient time is allowed even if the largest imaginary part of the spectrum is negative.

Physically, the transient time depends on the speed of propagating information, or the Lieb-Robinson Bounds \cite{Gong-18}. 
In a one-dimensional system with the non-Hermitian skin effect, the transient time is given by the time for the information to propagate from one end to the other end, which is proportional to the system size \cite{Gong-18}.
The corresponding ``quasi-eigenvalue" is in the $\epsilon$-psudospectrum with $\epsilon$ that is exponentially small with respect to the system size \cite{Gong-18}.
In the correspondence between non-Hermitian and Hermitian topology, the pseudospectrum under the non-Hermitian skin effect is related to the quasi-zero modes of a topological insulator \cite{Okuma-Sato-20}, as emphasized in Sect. \ref{pseudospectrum}.
In this sense, the transient dynamics under the non-Hermitian skin effect is related to the quasi-zero modes of a Hermitian topological insulator.
For general non-normal matrices, the transient time can be roughly estimated by the non-normal energy-time uncertainty relation defined for a pseudospectrum \cite{Okuma-Nakagawa}.

Which is physically more important, the spectrum or the pseudospectrum? The answer depends on the situation. 
If one is interested in the stability/instability under linear fluctuations, the stability/instability is judged by the negativity/positivity of the largest imaginary part of the spectrum.
If nonlinearity is not negligible, on the other hand, the combination of transient amplification and nonlinearity can lead to instability even if the largest imaginary part of the spectrum is negative \cite{Trefethen}. In addition to classical systems \cite{Trefethen, Asllani-2018}, the mathematics of stability/instability is relevant in quantum systems described by the bosonic Bogoliubov-de Gennes equation \cite{Mc-18,Flynn-20,Flynn-21,BdG-non-bloch,okuma2022boundary}.
In addition to the stability analysis, the relaxation process to the non-equilibrium steady state in the Lindblad equation is related to non-normality.
Several studies pointed out that the relaxation time is not always described by the spectral gap \cite{Okuma-Sato-20,haga-20,mori-20}. In our view, the origin of the mismatch is nothing but the pseudospectral nature of non-normal matrices.

Although the above discussion is limited to the Schor\"{o}dinger dynamics, we note that non-Hermitian spectral phenomena can appear in more generic situations.
For example, Ref. \cite{zou2021observation} detected the skin corner modes with a small imaginary part in electric circuits in which the relevant eigenvalues are not given by the largest imaginary part but by the smallest absolute value.
We hope that this review inspires readers to consider applications of non-Hermitian matrices to condensed matter physics.

\section*{ACKNOWLEDGMENTS}
The authors acknowledges Kenta Esaki, Kazuki Hasebe, Mahito Kohmoto, Kohei Kawabata, Ken Shiozaki, Masahito Ueda, Takumi Bessho,  Sayed Ali Akbar Ghorashi, Tianhe Li, Taylor L. Hughes and Yuya O. Nakagawa for collaborations. 
This work was supported by JST CREST Grant
No. JPMJCR19T2, JST ERATO-FS Grant No. JPMJER2105, and KAKENHI Grants Nos.JP20H00131 and JP20K14373.

\bibliography{NH-topo}

\begin{thebibliography}{147}
\expandafter\ifx\csname natexlab\endcsname\relax\def\natexlab#1{#1}\fi

\bibitem{Kane-review}
Hasan MZ, Kane CL. 2010.
\textit{Rev. Mod. Phys.} 82:3045

\bibitem{Zhang-review}
Qi XL, Zhang SC. 2011.
\textit{{Rev. Mod. Phys.}} 83:1057

\bibitem{Hatsugai-93}
Hatsugai Y. 1993.
\textit{Phys. Rev. Lett.} 71(22):3697--3700

\bibitem{Trefethen}
Trefethen LN, Embree M. 2005.
{Spectra and Pseudospectra}.
Princeton, NJ: Princeton University Press

\bibitem{Bender-98}
Bender CM, Boettcher S. 1998.
\textit{Phys. Rev. Lett.} 80:5243

\bibitem{Bender-02}
Bender CM, Brody DC, Jones HF. 2002.
\textit{{Phys. Rev. Lett.}} 89:270401

\bibitem{Bender-review}
Bender CM. 2007.
\textit{Rep. Prog. Phys.} 70:947

\bibitem{Konotop-review}
Konotop VV, Yang J, Zezyulin DA. 2016.
\textit{Rev. Mod. Phys.} 88:035002

\bibitem{Christodoulides-review}
El-Ganainy R, Makris KG, Khajavikhan M, Musslimani ZH, Rotter S,
  Christodoulides DN. 2018.
\textit{Nat. Phys.} 14:11

\bibitem{ashida-gong-20}
Ashida Y, Gong Z, Ueda M. 2020.
\textit{Advances in Physics} 69:249--435

\bibitem{Rudner-09}
Rudner MS, Levitov LS. 2009.
\textit{Phys. Rev. Lett.} 102(6):065703

\bibitem{Hu-11}
Hu YC, Hughes TL. 2011.
\textit{Phys. Rev. B} 84:153101

\bibitem{Esaki-11}
Esaki K, Sato M, Hasebe K, Kohmoto M. 2011.
\textit{Phys. Rev. B} 84:205128

\bibitem{Schomerus-13}
Schomerus H. 2013.
\textit{Opt. Lett.} 38:1912

\bibitem{Shen-18}
Shen H, Zhen B, Fu L. 2018.
\textit{Phys. Rev. Lett.} 120:146402

\bibitem{YW-18-SSH}
Yao S, Wang Z. 2018.
\textit{Phys. Rev. Lett.} 121:086803

\bibitem{Kunst-18}
Kunst FK, Edvardsson E, Budich JC, Bergholtz EJ. 2018.
\textit{Phys. Rev. Lett.} 121:026808

\bibitem{Hatano-Nelson-96}
Hatano N, Nelson DR. 1996.
\textit{Phys. Rev. Lett.} 77:570

\bibitem{Hatano-Nelson-97}
Hatano N, Nelson DR. 1997.
\textit{Phys. Rev. B} 56:8651

\bibitem{Hatano98}
Hatano N. 1998.
\textit{Physica A} 254:317

\bibitem{Schmidt-Spitzer-60}
Schmidt P, Spitzer F. 1960.
\textit{Math. Scand.} 8:15

\bibitem{YSW-18-Chern}
Yao S, Song F, Wang Z. 2018.
\textit{{Phys. Rev. Lett.}} 121:136802

\bibitem{YM-19}
Yokomizo K, Murakami S. 2019.
\textit{Phys. Rev. Lett.} 123:066404

\bibitem{yokomizo2020non}
Yokomizo K, Murakami S. 2020.
\textit{Progress of Theoretical and Experimental Physics} 2020(12):12A102

\bibitem{Yang-19}
Yang Z, Zhang K, Fang C, Hu J. 2020.
\textit{Phys. Rev. Lett.} 125:226402

\bibitem{KOS-20}
Kawabata K, Okuma N, Sato M. 2020.
\textit{Phys. Rev. B} 101(19):195147

\bibitem{Okuma-correlated}
Okuma N, Sato M. 2021{\natexlab{a}}.
\textit{Phys. Rev. Lett.} 126:176601

\bibitem{Lee-16}
Lee TE. 2016.
\textit{Phys. Rev. Lett.} 116:133903

\bibitem{MartinezAlvarez-18}
Martinez~Alvarez VM, Barrios~Vargas JE, Foa~Torres LEF. 2018.
\textit{Phys. Rev. B} 97:121401(R)

\bibitem{Xiong-2018}
Xiong Y. 2018.
\textit{Journal of Physics Communications} 2(3):035043

\bibitem{SSH-79}
Su WP, Schrieffer JR, Heeger AJ. 1979.
\textit{Phys. Rev. Lett.} 42:1698

\bibitem{song-real-space-19}
Song F, Yao S, Wang Z. 2019.
\textit{Phys. Rev. Lett.} 123(24):246801

\bibitem{KSUS-19}
Kawabata K, Shiozaki K, Ueda M, Sato M. 2019{\natexlab{a}}.
\textit{Phys. Rev. X} 9:041015

\bibitem{bogoliubov1947theory}
Bogoliubov N. 1947.
\textit{J. Phys} 11(1):23

\bibitem{Altland-Simons}
Altland A, Simons BD. 2010.
Condensed matter field theory.
Cambridge university press

\bibitem{Colpa-78}
Colpa J. 1978.
\textit{Physica A: Statistical Mechanics and its Applications} 93(3-4):327--353

\bibitem{kawaguchi2012spinor}
Kawaguchi Y, Ueda M. 2012.
\textit{Physics Reports} 520(5):253--381

\bibitem{Shindou-13}
Shindou R, Matsumoto R, Murakami S, Ohe Ji. 2013.
\textit{Phys. Rev. B} 87(17):174427

\bibitem{BdGsym}
Lieu S. 2018.
\textit{Phys. Rev. B} 98(11):115135

\bibitem{topo-magnon}
McClarty PA. 2022.
\textit{Annual Review of Condensed Matter Physics} 13(1):null

\bibitem{Ghatak-2019}
Ghatak A, Das T. 2019.
\textit{Journal of Physics: Condensed Matter} 31:263001

\bibitem{OKSS-20}
Okuma N, Kawabata K, Shiozaki K, Sato M. 2020.
\textit{Phys. Rev. Lett.} 124:086801

\bibitem{Zhang-19}
Zhang K, Yang Z, Fang C. 2020.
\textit{Phys. Rev. Lett.} 125:126402

\bibitem{Gong-18}
Gong Z, Ashida Y, Kawabata K, Takasan K, Higashikawa S, Ueda M. 2018.
\textit{Phys. Rev. X} 8:031079

\bibitem{Lee-Thomale-19}
Lee CH, Thomale R. 2019.
\textit{Phys. Rev. B} 99:201103(R)

\bibitem{Borgnia-19}
Borgnia DS, Kruchkov AJ, Slager RJ. 2020.
\textit{Phys. Rev. Lett.} 124:056802

\bibitem{Bottcher}
B\"ottcher A, Grudsky SM. 2005.
{Spectral Properties of Banded Toeplitz Matrices}.
Philadelphia: SIAM

\bibitem{lein2020choosing}
Lein M. 2020.
\textit{arXiv preprint arXiv:2010.09261}

\bibitem{roy-floquet-classification}
Roy R, Harper F. 2017.
\textit{Phys. Rev. B} 96(15):155118

\bibitem{Altland}
Altland A, Zirnbauer MR. 1997.
\textit{Phys. Rev. B} 55(2):1142--1161

\bibitem{Kawabata-19}
Kawabata K, Higashikawa S, Gong Z, Ashida Y, Ueda M. 2019{\natexlab{b}}.
\textit{Nat. Commun.} 10:297

\bibitem{Okuma-Sato-20}
Okuma N, Sato M. 2020.
\textit{Phys. Rev. B} 102(1):014203

\bibitem{Bauer-Fike}
Bauer FL, Fike CT. 1960.
\textit{Numer. Math.} 2:137

\bibitem{Golub-Van_Loan}
Golub G, Loan C. 1996.
Matrix computations (3rd ed.
The John Hopkins University Press

\bibitem{Budich-20}
Budich JC, Bergholtz EJ. 2020.
\textit{Phys. Rev. Lett.} 125(18):180403

\bibitem{Henrici}
P.Henrici. 1962.
\textit{Numer. Math.} 4:24

\bibitem{Nakai-Okuma-Sato}
Nakai Y, Okuma N, Sato M. 2022.
\textit{in preparation}

\bibitem{senthil2015symmetry}
Senthil T. 2015.
\textit{Annu. Rev. Condens. Matter Phys.} 6(1):299--324

\bibitem{Bernard-LeClair-02}
Bernard D, LeClair A. 2002.
``a classification of non-hermitian random matrices," in {\it statistical field
  theories} edited by a. cappelli and g. mussardo (springer, dordrecht), pp.
  207-214.

\bibitem{Yoshida-dagger-symmetry}
Yoshida T, Peters R, Kawakami N, Hatsugai Y. 2020.
\textit{Progress of Theoretical and Experimental Physics} 2020(12)

\bibitem{Lieu-19}
Lieu S, McGinley M, Cooper NR. 2020.
\textit{Phys. Rev. Lett.} 124:040401

\bibitem{teo-kane}
Teo JCY, Kane CL. 2010.
\textit{Phys. Rev. B} 82(11):115120

\bibitem{Sato-11}
Sato M, Hasebe K, Esaki K, Kohmoto M. 2012.
\textit{Prog. Theor. Phys.} 127:937

\bibitem{Kane-Mele-05-QSH}
Kane CL, Mele EJ. 2005.
\textit{Phys. Rev. Lett.} 95:226801

\bibitem{Okuma-19}
Okuma N, Sato M. 2019.
\textit{Phys. Rev. Lett.} 123:097701

\bibitem{Imura-12}
Imura KI, Yoshimura Y, Takane Y, Fukui T. 2012.
\textit{Phys. Rev. B} 86(23):235119

\bibitem{Parente-11}
Parente V, Lucignano P, Vitale P, Tagliacozzo A, Guinea F. 2011.
\textit{Phys. Rev. B} 83(7):075424

\bibitem{DLee-09}
Lee DH. 2009.
\textit{Phys. Rev. Lett.} 103(19):196804

\bibitem{Terrier-Kunst-20}
Terrier F, Kunst FK. 2020.
\textit{Phys. Rev. Research} 2(2):023364

\bibitem{Schnyder-08}
Schnyder AP, Ryu S, Furusaki A, Ludwig AWW. 2008.
\textit{Phys. Rev. B} 78:195125

\bibitem{Kitaev-09}
Kitaev A. 2009.
\textit{AIP Conf. Proc.} 1134:22

\bibitem{Ryu-10}
Ryu S, Schnyder AP, Furusaki A, Ludwig AWW. 2010.
\textit{New J. Phys.} 12:065010

\bibitem{Schnyder-Ryu-review}
Chiu CK, Teo JCY, Schnyder AP, Ryu S. 2016.
\textit{Rev. Mod. Phys.} 88:035005

\bibitem{ZL-19}
Zhou H, Lee JY. 2019.
\textit{Phys. Rev. B} 99:235112

\bibitem{Okuma-Sato-21}
Okuma N, Sato M. 2021{\natexlab{b}}.
\textit{Phys. Rev. B} 103(8):085428

\bibitem{Lee-Vishwanath-19}
Lee JY, Ahn J, Zhou H, Vishwanath A. 2019.
\textit{Phys. Rev. Lett.} 123:206404

\bibitem{Bessho-Sato-20}
Bessho T, Sato M. 2021.
\textit{Phys. Rev. Lett.} 127:196404

\bibitem{Rubakov}
Rubakov V. 1982.
\textit{Nucl. Phys. B} 203:311--348

\bibitem{callan}
Callan CG. 1982.
\textit{Phys. Rev. D} 25(8):2141--2146

\bibitem{fukushima2008chiral}
Fukushima K, Kharzeev DE, Warringa HJ. 2008.
\textit{Physical Review D} 78(7):074033

\bibitem{KBS-19}
Kawabata K, Bessho T, Sato M. 2019.
\textit{Phys. Rev. Lett.} 123:066405

\bibitem{stab-of-fermi-surface}
Ho\ifmmode~\check{r}\else \v{r}\fi{}ava P. 2005.
\textit{Phys. Rev. Lett.} 95(1):016405

\bibitem{matsuura2013protected}
Matsuura S, Chang PY, Schnyder AP, Ryu S. 2013.
\textit{New Journal of Physics} 15(6):065001

\bibitem{Zhao-PRL-13}
Zhao YX, Wang ZD. 2013.
\textit{Phys. Rev. Lett.} 110(24):240404

\bibitem{Zhao-PRB-14}
Zhao YX, Wang ZD. 2014.
\textit{Phys. Rev. B} 89(7):075111

\bibitem{Zhao-PRL-16}
Zhao YX, Schnyder AP, Wang ZD. 2016.
\textit{Phys. Rev. Lett.} 116(15):156402

\bibitem{Zhao-PRL-17}
Zhao YX, Lu Y. 2017.
\textit{Phys. Rev. Lett.} 118(5):056401

\bibitem{Morimoto-semimetal}
Morimoto T, Furusaki A. 2014.
\textit{Phys. Rev. B} 89(23):235127

\bibitem{Kobayashi-14}
Kobayashi S, Shiozaki K, Tanaka Y, Sato M. 2014.
\textit{Phys. Rev. B} 90(2):024516

\bibitem{Shiozaki-Sato-14}
Shiozaki K, Sato M. 2014.
\textit{Phys. Rev. B} 90(16):165114

\bibitem{CKC-14}
Chiu CK, Schnyder AP. 2014.
\textit{Phys. Rev. B} 90(20):205136

\bibitem{kato2013perturbation}
Kato T. 2013.
Perturbation theory for linear operators.
vol. 132.
Springer Science \& Business Media

\bibitem{berry2004physics}
Berry MV. 2004.
\textit{Czechoslovak journal of physics} 54(10):1039--1047

\bibitem{heiss2012physics}
Heiss W. 2012.
\textit{Journal of Physics A: Mathematical and Theoretical} 45(44):444016

\bibitem{Kozii-17}
Kozii V, Fu L. 2017.
\textit{arXiv preprint, arXiv:1708.05841}

\bibitem{zhou-18}
Zhou H, Peng C, Yoon Y, Hsu CW, Nelson KA, et~al. 2018.
\textit{Science} 359:1009

\bibitem{Papaj-19}
Papaj M, Isobe H, Fu L. 2019.
\textit{Phys. Rev. B} 99(20):201107

\bibitem{Bergholtz-review}
Bergholtz EJ, Budich JC, Kunst FK. 2021.
\textit{Rev. Mod. Phys.} 93(1):015005

\bibitem{Zhou-19}
Zhou H, Lee JY, Liu S, Zhen B. 2019.
\textit{Optica} 6:190

\bibitem{Zyuzin-18}
Zyuzin AA, Zyuzin AY. 2018.
\textit{Phys. Rev. B} 97:041203(R)

\bibitem{Yoshida-18}
Yoshida T, Peters R, Kawakami N. 2018.
\textit{Phys. Rev. B} 98:035141

\bibitem{Yoshida-19}
Yoshida T, Peters R, Kawakami N, Hatsugai Y. 2019.
\textit{Phys. Rev. B} 99:121101(R)

\bibitem{Bergholtz-19}
Bergholtz EJ, Budich JC. 2019.
\textit{Phys. Rev. Research} 1:012003(R)

\bibitem{Kimura-19}
Kimura K, Yoshida T, Kawakami N. 2019.
\textit{{Phys. Rev. B}} 100:115124

\bibitem{Zhen-15}
Zhen B, Hsu CW, Igarashi Y, Lu L, Kaminer I, et~al. 2015.
\textit{Nature} 525:354

\bibitem{Xu-17}
Xu Y, Wang ST, Duan LM. 2017.
\textit{Phys. Rev. Lett.} 118:045701

\bibitem{Okugawa-19}
Okugawa R, Yokoyama T. 2019.
\textit{Phys. Rev. B} 99:041202(R)

\bibitem{Budich-19}
Budich JC, Carlstr\"om J, Kunst FK, Bergholtz EJ. 2019.
\textit{Phys. Rev. B} 99:041406(R)

\bibitem{Cerjan-19}
Cerjan A, Huang S, Chen KP, Chong Y, Rechtsman MC. 2019.
\textit{Nat. Photon.} 13:623

\bibitem{Brandenbourger-19-skin-exp}
Brandenbourger M, Locsin X, E.~Lerner CC. 2019.
\textit{Nat. Commun.} 10:4608

\bibitem{Ghatak-19-skin-exp}
Ghatak A, Brandenbourger M, Van~Wezel J, Coulais C. 2020.
\textit{Proceedings of the National Academy of Sciences} 117(47):29561--29568

\bibitem{Helbig-19-skin-exp}
Helbig T, Hofmann T, Imhof S, Abdelghany M, Kiessling T, et~al. 2020.
\textit{Nature Physics} 16(7):747--750

\bibitem{Hofmann-19-skin-exp}
Hofmann T, Helbig T, Schindler F, Salgo N, Brzezi{\'n}ska M, et~al. 2020.
\textit{Physical Review Research} 2(2):023265

\bibitem{Yi-Yang-20}
Yi Y, Yang Z. 2020.
\textit{Phys. Rev. Lett.} 125(18):186802

\bibitem{Xiao-19-skin-exp}
Xiao L, Deng T, Wang K, Zhu G, Wang Z, et~al. 2020.
\textit{Nature Physics} 16(7):761--766

\bibitem{Sato-Ando}
Sato M, Ando Y. 2017.
\textit{Reports on Progress in Physics} 80(7):076501

\bibitem{Sato-Fujimoto}
Sato M, Takahashi Y, Fujimoto S. 2009.
\textit{Phys. Rev. Lett.} 103(2):020401

\bibitem{Lutchyn}
Lutchyn RM, Sau JD, Das~Sarma S. 2010.
\textit{Phys. Rev. Lett.} 105(7):077001

\bibitem{O-R-vO}
Oreg Y, Refael G, von Oppen F. 2010.
\textit{Phys. Rev. Lett.} 105(17):177002

\bibitem{Lee-Li-Gong-19}
Lee CH, Li L, Gong J. 2019.
\textit{Phys. Rev. Lett.} 123:016805

\bibitem{zou2021observation}
Zou D, Chen T, He W, Bao J, Lee CH, et~al. 2021.
\textit{Nature Communications} 12(1):1--11

\bibitem{Okugawa-takahashi-yokomizo}
Okugawa R, Takahashi R, Yokomizo K. 2020.
\textit{Phys. Rev. B} 102(24):241202

\bibitem{KSS-20}
Kawabata K, Sato M, Shiozaki K. 2020.
\textit{Phys. Rev. B} 102(20):205118

\bibitem{Fu-Hu-Wan-21}
Fu Y, Hu J, Wan S. 2021.
\textit{Phys. Rev. B} 103(4):045420

\bibitem{benalcazar2017quantized}
Benalcazar WA, Bernevig BA, Hughes TL. 2017{\natexlab{a}}.
\textit{Science} 357(6346):61--66

\bibitem{benalcazar2017electric}
Benalcazar WA, Bernevig BA, Hughes TL. 2017{\natexlab{b}}.
\textit{Physical Review B} 96(24):245115

\bibitem{langbehn2017reflection}
Langbehn J, Peng Y, Trifunovic L, von Oppen F, Brouwer PW. 2017.
\textit{Physical review letters} 119(24):246401

\bibitem{song2017d}
Song Z, Fang Z, Fang C. 2017.
\textit{Physical review letters} 119(24):246402

\bibitem{Liu-19}
Liu T, Zhang YR, Ai Q, Gong Z, Kawabata K, et~al. 2019.
\textit{Phys. Rev. Lett.} 122:076801

\bibitem{Edvardsson-19}
Edvardsson E, Kunst FK, Bergholtz EJ. 2019.
\textit{Phys. Rev. B} 99:081302(R)

\bibitem{GLSH-21}
Ghorashi SAA, Li T, Sato M, Hughes TL. 2021.
\textit{Phys. Rev. B} 104(16):L161116

\bibitem{GLS-21}
Ghorashi SAA, Li T, Sato M. 2021.
\textit{Phys. Rev. B} 104(16):L161117

\bibitem{wu2020observation}
Wu J, Huang X, Lu J, Wu Y, Deng W, et~al. 2020.
\textit{Physical Review B} 102(10):104109

\bibitem{zhang2021observation}
Zhang X, Tian Y, Jiang JH, Lu MH, Chen YF. 2021.
\textit{Nature communications} 12(1):1--8

\bibitem{Lindblad}
Lindblad G. 1976.
\textit{Commun. Math. Phys.} 48:119

\bibitem{Prosen-2008}
Prosen T. 2008.
\textit{New Journal of Physics} 10(4):043026

\bibitem{Prosen-2010}
Prosen T. 2010.
\textit{Journal of Statistical Mechanics: Theory and Experiment}
  2010(07):P07020

\bibitem{Daley-14}
Daley AJ. 2014.
\textit{Advances in Physics} 63(2):77--149

\bibitem{Asllani-2018}
Asllani M, Lambiotte R, Carletti T. 2018.
\textit{Science Advances} 4(12):eaau9403

\bibitem{Okuma-Nakagawa}
Okuma N, Nakagawa YO. 2022.
\textit{Phys. Rev. B} 105(5):054304

\bibitem{Mc-18}
{A. McDonald, T. Pereg-Barnea, and A. A. Clerk}. 2018.
\textit{Phys. Rev. X} 8(4):041031

\bibitem{Flynn-20}
{Flynn, Vincent P and Cobanera, Emilio and Viola, Lorenza}. 2020.
\textit{New Journal of Physics} 22(8):083004

\bibitem{Flynn-21}
{Flynn, Vincent P. and Cobanera, Emilio and Viola, Lorenza}. 2021.
\textit{Phys. Rev. Lett.} 127(24):245701

\bibitem{BdG-non-bloch}
Yokomizo K, Murakami S. 2021.
\textit{Phys. Rev. B} 103(16):165123

\bibitem{okuma2022boundary}
Okuma N. 2022.
\textit{arXiv preprint arXiv:2202.07684}

\bibitem{haga-20}
Haga T, Nakagawa M, Hamazaki R, Ueda M. 2021.
\textit{Phys. Rev. Lett.} 127(7):070402

\bibitem{mori-20}
Mori T, Shirai T. 2020.
\textit{Phys. Rev. Lett.} 125(23):230604

\end{thebibliography}


\begin{thebibliography}{00}

\bibitem{Trouve1995a}
Trouv\'{e} A. 1995. {\it An approach of pattern recognition through
infinite dimensional group action.} Rep. LMENS-95-9, Lab. Math. l'Ecole Norm. Superieure, Paris  

\bibitem{Christensen1996}
Christensen G, Miller MI, Rabbit RD. 1995.  {\it IEEE Trans. Med. Imaging}
5(10):1435--47

\bibitem{Grenander1998}
Grenander U, Miller MI. 1998.  {\it Q. Appl. Math.} 56:617--94

\bibitem{Dupuis1998}%4
Dupuis P, Grenander U, Miller MI. 1998. {\it Q. Appl. Math.}
56:587--600

\bibitem{Miller-Younes-2001}
Miller MI, Younes L. 2001.  {\it Int. J. Comput. Vis.}
41:61--84

\bibitem{Toga2001}
Toga A, Thompson PM. 2001.  {\it Anat. Rec.}
265:37--53



 

\end{thebibliography}
\bibliographystyle{ar-style4.bst} 

\end{document}